\documentclass[superscriptaddress,10pt,twocolumn,aps,prl]{revtex4-2}
\usepackage{amsmath}
\usepackage{graphicx}
\usepackage{hyperref}
\hypersetup{colorlinks,allcolors=blue}
\begin{document}

\title{Charge density waves in infinite-layer NdNiO$_2$ nickelates}

\author{Charles C. Tam}
    \thanks{These authors contributed equally}
    \affiliation{Diamond Light Source, Harwell Campus, Didcot OX11 0DE, United Kingdom.}
    \affiliation{H. H. Wills Physics Laboratory, University of Bristol, Bristol BS8 1TL, United Kingdom.}

\author{Jaewon Choi}
    \thanks{These authors contributed equally}
    \affiliation{Diamond Light Source, Harwell Campus, Didcot OX11 0DE, United Kingdom.}
    
\author{Xiang Ding}
    \thanks{These authors contributed equally}
    \affiliation{School of Physics, University of Electronic Science and Technology of China, Chengdu, 610054, China}

\author{Stefano Agrestini}
    \affiliation{Diamond Light Source, Harwell Campus, Didcot OX11 0DE, United Kingdom.}

\author{Abhishek Nag}
    \affiliation{Diamond Light Source, Harwell Campus, Didcot OX11 0DE, United Kingdom.}
    \affiliation{Laboratory for Non-linear Optics, Paul Scherrer Institut, CH-5232 Villigen, Switzerland}
    
\author{Mei Wu}
    \affiliation{International Center for Quantum Materials and Electron Microscopy Laboratory, School of Physics, Peking University, Beijing 100871, China}

\author{Bing Huang}
    \affiliation{Beijing Computational Science Research Center, Beijing 100193, China}

\author{Huiqian Luo}
    \affiliation{Beijing National Laboratory for Condensed Matter Physics, Institute of Physics, Chinese Academy of Sciences, Beijing 100190}
    \affiliation{Songshan Lake Materials Laboratory, Dongguan, Guangdong 523808, China}
    
\author{Peng Gao}
    \affiliation{International Center for Quantum Materials and Electron Microscopy Laboratory, School of Physics, Peking University, Beijing 100871, China}

\author{Mirian Garc\'ia-Fern\'andez}
    \affiliation{Diamond Light Source, Harwell Campus, Didcot OX11 0DE, United Kingdom.}

\author{Liang Qiao}
    \email{liang.qiao@uestc.edu.cn}
    \affiliation{School of Physics, University of Electronic Science and Technology of China, Chengdu, 610054, China}

\author{Ke-Jin Zhou}
    \email{kejin.zhou@diamond.ac.uk}
    \affiliation{Diamond Light Source, Harwell Campus, Didcot OX11 0DE, United Kingdom.}


\begin{abstract}
In materials science, much effort has been devoted to reproduce superconductivity in chemical compositions analogous to cuprate superconductors since their discovery over thirty years ago. This approach was recently successful in realising superconductivity in infinite-layer nickelates~\cite{li2019superconductivity,li2020superconducting,zeng2020phase,osada2020phase,osada2021nickelate,zeng2021superconductivity}. Although differing from cuprates in electronic and magnetic properties, strong Coulomb interactions suggest infinite-layer nickelates have a propensity to various symmetry-breaking orders that populate the cuprates~\cite{hepting2020electronic, lu2021magnetic,tranquada1995evidence,ghiringhelli2012long}. Here we report the observation of charge density waves (CDWs) in infinite-layer NdNiO$_2$ films using Ni-$L_3$ resonant x-ray scattering. Remarkably, CDWs form in Nd 5$d$ and Ni 3$d$ orbitals at the same commensurate wavevector $(0.333, 0)\;r.l.u.$, with non-negligible out-of-plane dependence, and an in-plane correlation length up to $\sim$ 60 \AA. Spectroscopic studies reveal a strong connection between CDWs and the Nd 5$d$ - Ni 3$d$ orbital hybridisation. Upon entering the superconducting state at 20\% Sr doping, the CDWs disappear. Our work demonstrates the existence of CDWs in infinite-layer nickelates with a multi-orbital character distinct from cuprates, which establishes their low-energy physics.
\end{abstract}

\maketitle

The realisation of superconducting infinite-layer nickelates marks the latest milestone in the field of high-temperature superconductivity research~\cite{li2019superconductivity,li2020superconducting,zeng2020phase,osada2020phase,osada2021nickelate,zeng2021superconductivity}. Being iso-structural to CaCuO$_2$, infinite-layer nickelates contain quasi-two-dimensional (2D) NiO$_2$ layers, nominal $3d^9$ Ni$^{1+}$ ions with spin $S = 1/2$, and an active $d_{x^2-y^2}$ orbital near the Fermi level, analogous to the cuprate family of high-temperature superconductors~\cite{lee2004infinite}. However, X-ray absorption (XAS) and electron energy loss spectroscopies have shown their electronic structure is closer to the Mott-Hubbard than the charge-transfer regime, distinct from the cuprates~\cite{hepting2020electronic,goodge2021doping}. Another distinctive feature when comparing to cuprates is that the three-dimensional (3D) itinerant $5d$ bands of the rare-earth ions ($R$ in $R$NiO$_2$) are predicted to hybridise with the localised 2D Ni-O bands~\cite{lee2004infinite,hepting2020electronic,been2021electronic}. This strong hybridisation is corroborated by resonant inelastic X-ray scattering (RIXS) where Nd $5d$ - Ni $3d$ hybridised states were observed~\cite{hepting2020electronic}, as well as a change in sign of the low-temperature Hall coefficient as a function of Sr doping, indicating the presence of two bands at the Fermi level~\cite{li2020superconducting,zeng2020phase}.

On the other hand, although the magnetism of infinite-layer nickelates is under debate~\cite{hayward2003synthesis,yi2021nmr}, Nd$_{1-x}$Sr$_x$NiO$_2$, grown on (and capped with) SrTiO$_3$ (STO), shows well defined and highly dispersive magnetic excitations, validating the existence of strong electron Coulomb interactions in infinite-layer nickelates. In particular, this puts nickelates in the proximity between strong antiferromagnetic (AFM) correlations and superconductivity~\cite{lu2021magnetic}. The strong electronic and AFM correlations are key ingredients that give rise to symmetry-breaking orders, such as spin density waves (SDWs), or charge density waves (CDWs), which are relevant to superconductivity in the cuprates~\cite{fradkin2015colloquium}. A natural question is whether these ordered states are also present in infinite-layer nickelates. In this study, we measured thin films of NdNiO$_2$ with Ni-$L_3$-edge XAS and RIXS and revealed the presence of CDWs. We show that there is a clear correlation between the CDWs and the hybridised Nd $5d$ - Ni $3d$ orbital, demonstrating the active participation of the rare-earth 5$d$ states in the low-energy physics of the nickelates.     

\begin{figure*}[!htb]
	\begin{center}
	\includegraphics[width=1\textwidth]{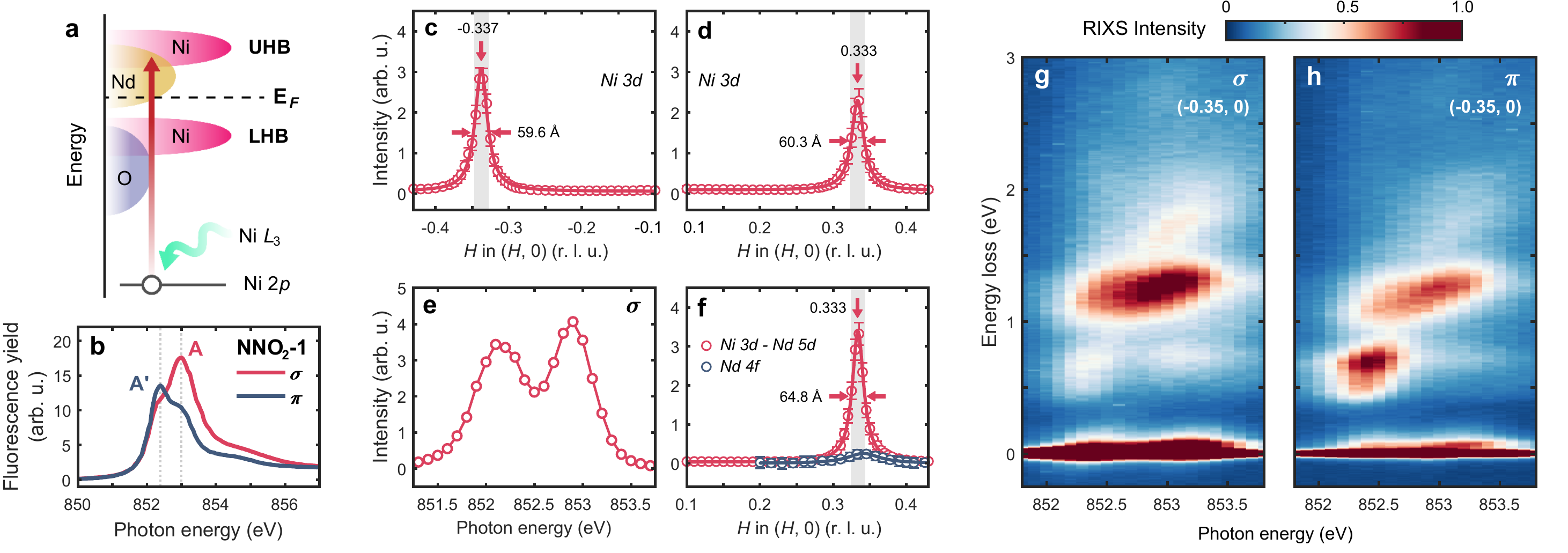}
		\caption{\textbf{CDWs in the parent NdNiO$_2$ thin film NNO$_2$-1.} \textbf{a,}
			Schematic electronic structure of the infinite-layer nickelates. \textbf{b,} Ni $L_3$ XAS of the parent NdNiO$_2$ film NNO$_2$-1. \textbf{c} and \textbf{d,} Integrated quasielastic peak intensity as a function of momentum transfer along the $(\pm H, 0)$ direction by exciting at the Ni 3$d$ resonance (A peak in \textbf{b}). The sign of the momentum transfer is defined in Methods (See also Supplementary Fig.~S8). Fitted peak centre values are $-0.337\pm0.002\;r.l.u.$ and $0.333\pm0.002\;r.l.u.$. Fitted correlation lengths are $59.6\pm1.2\;$\AA{} and $60.3\pm1.3\;$\AA{}. \textbf{e,} The resonant energy profile of the CDW by fixing the momentum transfer to $Q_\mathrm{CDW}$ = $(+0.333, 0)$. \textbf{f,} Integrated quasielastic peaks as a function of the momentum transfer along $(H, 0)$ direction by probing the Ni $3d$-Nd $5d$ hybridised state (A' peak in \textbf{b}) and Nd 4$f$ states at the Nd $M_5$ edge. Fitted peak centre is $0.333\pm0.002\;r.l.u.$ and fitted correlation length is $64.8\pm1.8\;$\AA{}. \textbf{g} and \textbf{h,} RIXS maps with photon energy varied across the Ni $L_3$-edge on NNO$_2$-1 at 20~K at $H=-0.35\;r.l.u.$ with $\sigma$ and $\pi$ incident x-ray polarisation, respectively.
		} \label{fig1}
	\end{center}
\end{figure*}

Nd$_{1-x}$Sr$_x$NiO$_2$ thin films (NdNiO$_2$ are 10 nm and Sr-doped Nd$_{1-x}$Sr$_x$NiO$_2$ is 15 nm) were prepared by pulsed laser deposition and subsequently topotactic reduction, similar to methods applied in Refs.~\cite{li2019superconductivity,zeng2020phase} but without STO capping layers (see Methods). We carried out  structural analysis using X-ray diffraction, X-ray reflectivity, reciprocal space mapping and rocking curve scans of the (002) diffraction peak, atomic force microscopy, and scanning transmission electron microscopy (Methods and Supplementary Note 1). All thin films show comparable crystalline quality with a vast majority of the square-planar phase and a minor presence of the Ruddlesden-Popper (RP) secondary phase. The level of inhomogeneity (3.7\% and 15.2\% for the bulk and the surface, respectively) was estimated from the analysis of the O $K$-XAS spectra (Supplementary Note 2). 

One of the reduced parent NdNiO$_2$ films, referred as 'NNO$_2$-1' hereafter, features a two-peak Ni $L_3$ XAS profile consistent to those in LaNiO$_2$ and NdNiO$_2$ reported recently~\cite{hepting2020electronic,rossi2020orbital}. The first peak (denoted A') reflects the transition to the Nd $5d$-Ni $3d$ hybridised states and the second peak (denoted A) detects the main absorption from $2p_{3/2}^6$3$d^9$ to $2p_{3/2}^5$3$d^{10}$ transition at the nominal Ni$^{1+}$ sites~(Figs.~\ref{fig1}a, b, and Supplementary Fig.~S8)~\cite{hepting2020electronic,rossi2020orbital}. The O $K$ XAS spectra also shows good consistency with other infinite-layer nickelates, in the bulk and the surface of the film, namely the hole-peak weight is significantly suppressed at the pre-edge owing to a much reduced O 2$p$-Ni $3d$ hybridisation comparing to pristine NdNiO$_3$ (Supplementary Note 4)~\cite{hepting2020electronic,goodge2021doping,chen2021electronic}.

To begin the search for symmetry-breaking orders, we performed momentum-dependent RIXS on NNO$_2$-1 ($\sigma$ polarisation is used throughout unless otherwise stated). By exciting at the Ni $L_3$ resonance peak (A in Fig.~\ref{fig1}b), quasielastic scattering peaks appear at $q_{\parallel} = - 0.337 \pm 0.002$ and $q_{\parallel} = 0.333 \pm 0.002\;r.l.u.$ along the primary $(H,0)$, \textit{i.e.}, the Ni-O bonding direction (Figs.~\ref{fig1}c and d). Scanning in the $(H, H)$ direction however did not reveal such peaks (Supplementary Fig.~S9). To clarify the origin of the scattering peak, we fixed $q_{\parallel}$ to +0.333\;$r.l.u.$ and swept the incident photon energy across the Ni $L_3$ absorption edge. Interestingly, a pronounced double-resonance profile is found coinciding with the two absorption peaks (Fig.~\ref{fig1}e). Photon energy scans at $q_{\parallel} = -0.35\;r.l.u.$ across the Ni $L_3$ reveal a much stronger signal with $\sigma$ than $\pi$ polarisation. Likewise, the quasielastic peak in the momentum space possesses the same polarisation dependence as charge excitations and opposite to that of the spin excitations in cuprates (Supplementary Figs.~S11 and S12)~\cite{ghiringhelli2012long}. In addition, momentum-dependent RIXS scans off the Ni $L_3$ resonance peak, taken at 840 eV, returned negligible quasielastic peaks (Supplementary Fig.~S9). The above results suggest the observed quasielastic scattering peak may be attributed to translational symmetry breaking induced by a charge density modulation. While there is potential contribution from a structural modulation or the defect phase to the quasielastic peak, we will continue to refer to the phenomena as CDW and discuss these possibilities later. Notably, the double resonance behaviour contrasts to the CDWs in cuprates where typically a singular resonance profile exists at the Cu $L_3$ edge~\cite{li2020multiorbital}.     
\begin{figure*}[!htb]
	\begin{center}
		\includegraphics[width=1\textwidth]{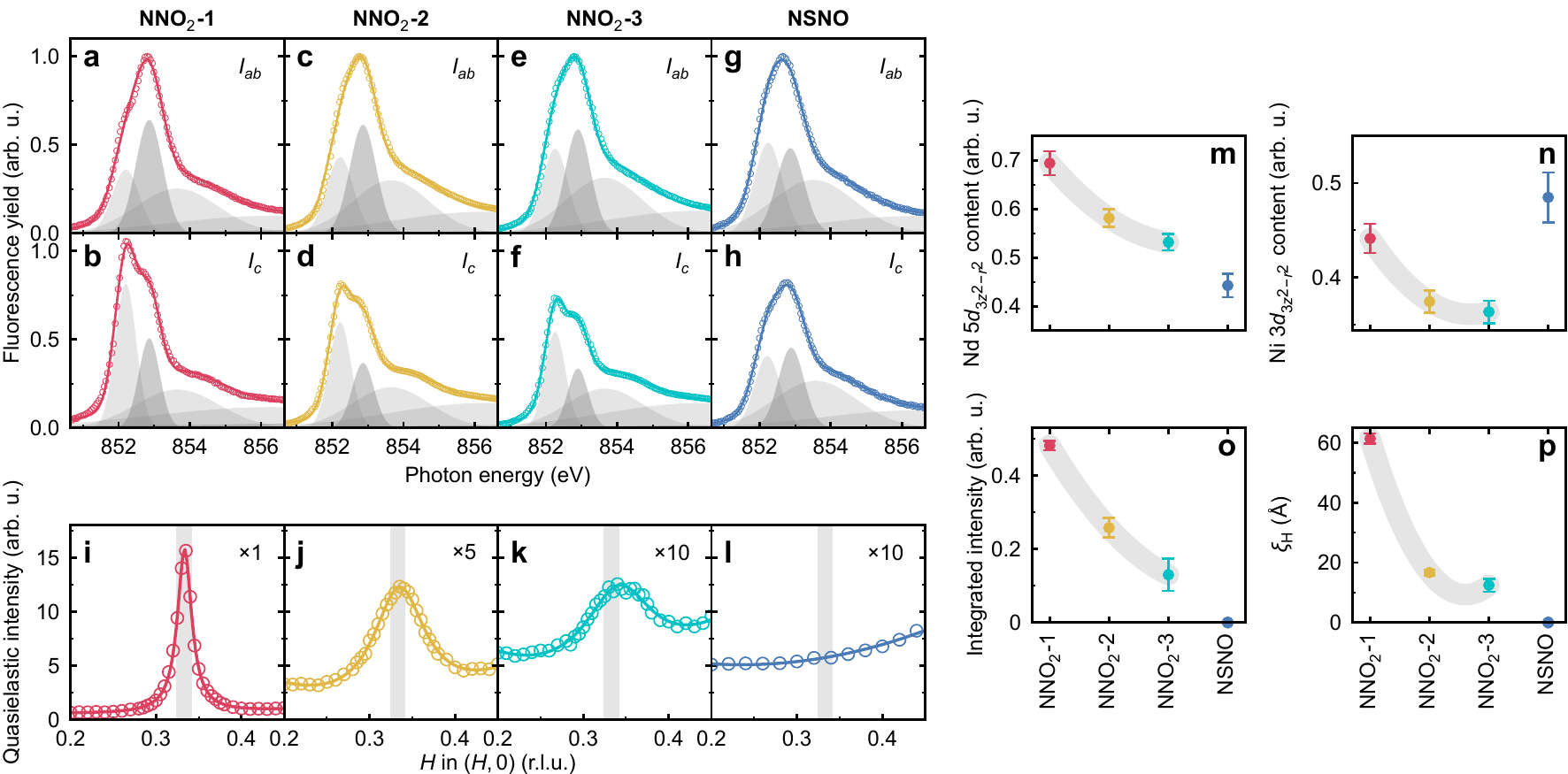}
		\caption{\textbf{Nd 5$d$-Ni 3$d$ orbital hybridisation and CDW in NdNiO$_2$ and superconducting Nd$_{0.8}$Sr$_{0.2}$NiO$_2$}. The XAS projected in (out of) the NiO$_2$ planes, I$_{ab}$ (I$_c$), of NNO$_2$-1, NNO$_2$-2, NNO$_2$-3, and NSNO are shown in \textbf{a} (\textbf{b}), \textbf{c} (\textbf{d}), \textbf{e} (\textbf{f}), \textbf{g} (\textbf{h}), respectively. The fit components that are visible in this energy range are plotted in different grey colors (See fitting details in Methods). The quasielastic peak intensity showing a CDW in the three NdNiO$_2$ films are plotted in \textbf{i}, \textbf{j}, \textbf{k}, with no evidence of CDW in NSNO (\textbf{l}). The Nd $5d_{3z^2-r^2}$ orbital content (\textbf{m}), the Ni $3d_{3z^2-r^2}$ orbital content (\textbf{n}), the CDW integrated intensity (\textbf{o}), and the CDW in-plane correlation length (\textbf{p}) show similar trends for the NdNiO$_2$ samples. Error bars are from least-squares fitting.}
		 \label{fig2}
	\end{center}
\end{figure*}

Density-functional theory studies on the infinite-layer nickelates uncovered sizable Ni $3d_{3z^2-r^2}$ mixing with the rare-earth $5d_{3z^2-r^2}$ and $5d_{xy}$ states, that leads to 3D Fermi surface pockets that slightly hole dope the half-filled Ni $3d_{x^2-y^2}$ band~\cite{lee2004infinite,sakakibara2020model,been2021electronic}. As a result, Ni $3d_{3z^2-r^2}$ weight appears near the Fermi level, although the Ni $3d_{3z^2-r^2}$ orbital is furthest away from the Ni $3d_{x^2-y^2}$ orbital in a simplified ligand field picture. The pre-peak in the Ni $L_3$ XAS spectra as well as the $\sim0.6$\;eV energy loss feature in RIXS acquired in both $\sigma$ and $\pi$ polarisations are signatures of the hybridised Nd $5d$-Ni $3d$ orbitals containing $5d_{3z^2-r^2}$- and $5d_{xy}$- symmetries unique in the infinite-layer nickelates (Fig.~\ref{fig1}b, g and h). We thus positioned the photon energy at the Ni-Nd hybridised states (A' in Fig.~\ref{fig1}b) and scanned along the $(H, 0)$ direction. Remarkably, a quasielastic scattering peak appears at the same wavevector as those at the Ni $3d$ resonance (Fig.~\ref{fig1}f). Moreover, the quasielastic CDW peaks at Nd and Ni resonances show a comparable half-width at half-maximum (HWHM) $\Gamma = 0.01\pm 0.002\;r.l.u.$ We define the in-plane correlation length $\xi_H$ =  1/$\Gamma$ which corresponds to $60.3\pm1.3$\;\AA{} (Figs.~\ref{fig1}c-f). The above results indicate that both Nd 5$d$ and Ni 3$d$ valence charges form density waves in NNO$_2$-1 propagating with the same period. A quasielastic scattering peak also appears at $q_{\parallel} = 0.340 \pm 0.004\;r.l.u.$ when exciting at the Nd $M_5$-edge (Fig.~\ref{fig1}f). The significantly weakened intensity suggests that the localised $4f$ states are much less involved in the valence band near the Fermi level~\cite{hepting2020electronic}.




To explore the properties of CDWs in infinite-layer NdNiO$_2$, we designed a series of NdNiO$_2$ thin films by tuning the annealing temperature (See Methods). Figs.~\ref{fig2}a-h summarise the Ni $L_3$ XAS of the two new NdNiO$_2$ (denoted as NNO$_2$-2, and NNO$_2$-3, hereafter) together with that of NNO$_2$-1. The two-peak structure is clearly present in all XAS spectra projected along the direction of the in-plane, $I_\text{ab}$, and the out-of- plane, $I_\text{c}$ (See Methods). Considering NdNiO$_2$ as a nominal $d^9$ system, we fitted the two peaks of the projected XAS spectra from which the Nd $5d$ and Ni $3d$ orbital occupancy can be extracted (See Methods). The Nd $5d_{3z^2-r^2}$ and Ni $3d_{3z^2-r^2}$ orbital content is thus defined as $I_\text{c}/\left(I_\text{c} + I_\text{ab}\right)$. Figs.~\ref{fig2}m and n show both the Nd 5$d_{3z^2-r^2}$ and Ni $3d_{3z^2-r^2}$ orbital contents decrease progressively, signalling reduced Nd-Ni hybridisation from NNO$_2$-1 to NNO$_2$-3. The same monotonic trend is found from the surface layers of the samples (Supplementary Fig.~S15). Conversely, the O $K$- XAS spectra are consistent across the three samples from the bulk and the surface layers (Supplementary Note 4). RIXS maps of NNO$_2$-2 and NNO$_2$-3 also manifest qualitatively the same $dd$ excitations as those in NNO$_2$-1 (Supplementary Note 3). The consistent bulk and  surface spectroscopic results among all NdNiO$_2$ films suggest the variation of the Nd-Ni hybridisation is a property of the majority square-planar phase rather than the minor RP defect phase. Figs.~\ref{fig2}i-l summarise the integrated quasielastic peak intensities as a function of $q_{\parallel}$ along the $(H, 0)$ direction, showing clearly CDWs in all NdNiO$_2$ samples. To make a quantitative assessment, we compare the fitted CDW intensity and in-plane correlation length. Figs.~\ref{fig2}o and p display monotonically decreasing CDW integrated intensity and the in-plane correlation length, respectively, from NNO$_2$-1 to NNO$_2$-3. A similar trend between CDWs and the $d_{3z^2-r^2}$ orbital content across three parent NdNiO$_2$
suggesting that CDWs may hold an intimate connection with the Nd-Ni hybridisation (Figs.~\ref{fig2}m and n). Noticeably, the CDWs in NdNiO$_2$ are commensurate, propagating along the Ni-O bonding direction, distinct to the charge order formed along the Ni-Ni bonding direction in both single-layer La$_{2-x}$Sr$_x$NiO$_4$ and trilayer La$_{4}$Ni$_3$O$_8$ nickelates~\cite{cheong1994charge,zhang2016stacked}.

We also studied a superconducting Nd$_{0.8}$Sr$_{0.2}$NiO$_2$ film, with $T_c = 10\;$K (referred as NSNO hereafter). XAS and RIXS measurements were conducted under the same experimental geometry. Whereas NSNO presents similar O $K$ XAS spectra compared to the parent samples (Supplementary Note 4), the Ni $L_3$ $I_\text{c}$ XAS changes noticeably in that the Nd resonance peak becomes much reduced. Comparing to the parent samples, the Nd $5d_{3z^2-r^2}$ orbital content decreases sharply, indicating the Nd-Ni hybridisation is further reduced in the superconducting sample (Figs.~\ref{fig2}m). This is in line with the fact that 20\% Sr doping depletes most of the itinerant Nd 5$d$ electron carriers~\cite{li2020superconducting,zeng2020phase}. The Ni $3d_{3z^2-r^2}$ orbital content increases, deviating from the trend in both the bulk and the surface (Figs.~\ref{fig2}n and Supplementary Fig.~S13), which suggests the orbital character becomes more isotropic than in NdNiO$_2$, which supports the picture that the Ni states get more $d^8$-like. Interestingly, no CDW signals were detected along either the $(H, 0)$ or $(H, H)$ directions (Fig.~\ref{fig2}l and Supplementary Fig.~S9). The results of the parent and the superconducting samples suggest that the Nd $5d$ hybridised orbital actively contributes to the CDW ordered states in the infinite-layer nickelates. The situation is disparate to cuprates where the conventional CDWs are normally hosted in the CuO$_2$ layers, not in the spacer-layers, and are quasi-two-dimensional~\cite{ghiringhelli2012long,comin2016resonant}. 

\begin{figure}[!htb]
	\begin{center}
		\includegraphics[width=.5\textwidth]{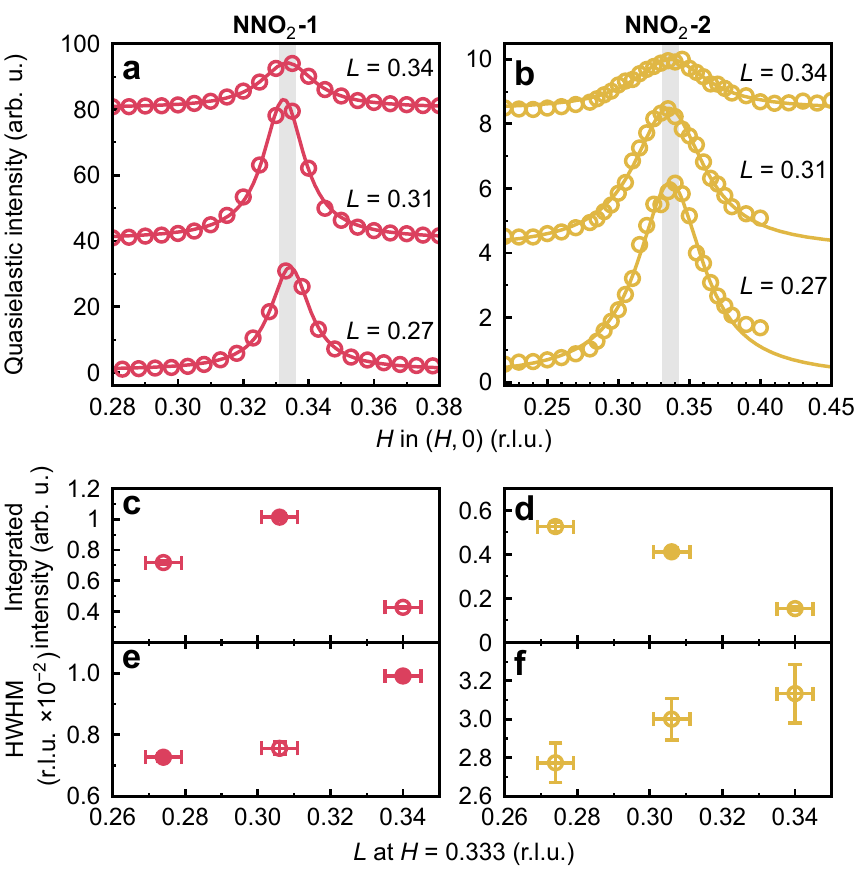}
		\caption{\textbf{$L$-dependence of CDW in NdNiO$_2$.} \textbf{a,} and \textbf{b,} Integrated quasielastic peaks and their Lorentzian fits along ($H$, 0) direction with each acquired at a different $\Omega$ value for NNO$_2$-1 and NNO$_2$-2, respectively. $\Omega$ was chosen such that the CDW scattering is peaked at $H=0.333$ with the corresponded $L$ value labeled in \textbf{a} and \textbf{b}. Measurements at successively higher $L$ values have been offset for clarity. \textbf{c,} and \textbf{d,} Integrated intensity of the CDW peak as a function of $L$ at $H=0.333$. \textbf{e,} and \textbf{f,} The CDW peak width (HWHM) as a function of $L$. Error bars of the intensity and width in \textbf{c}-\textbf{f} are from the least squares fitting. Horizontal error bars in \textbf{c}-\textbf{f} represent the momentum resolution at the Ni $L_3$ RIXS.}
		 \label{fig3}
	\end{center}
\end{figure}

To further explore the properties of CDWs in the infinite-layer nickelates, we examined the $L$-dependence. In Figs.~\ref{fig3}a and b are plotted a series of CDW scans peaked at $(0.333, 0, L)$ from both NNO$_2$-1 and NNO$_2$-2, where $L$ is changed discretely (See Methods). Notably, the integrated CDW intensity (Figs.~\ref{fig3}c and d) and peak width (Figs.~\ref{fig3}e and f) change significantly when $L$ is decreased from 0.34 to 0.27. Due to the limit of the Ni $L_3$ resonance energy and a relatively short $c$-axis lattice parameter, only a very small portion of the $L$ space is accessible. Nevertheless, the non-negligible $L$-dependence implies CDWs may have a three-dimensional nature. Further studies using hard x-ray scattering may shed light on the question of the dimensionality.     

\begin{figure*}[!htb]
	\begin{center}
		\includegraphics[width=\textwidth]{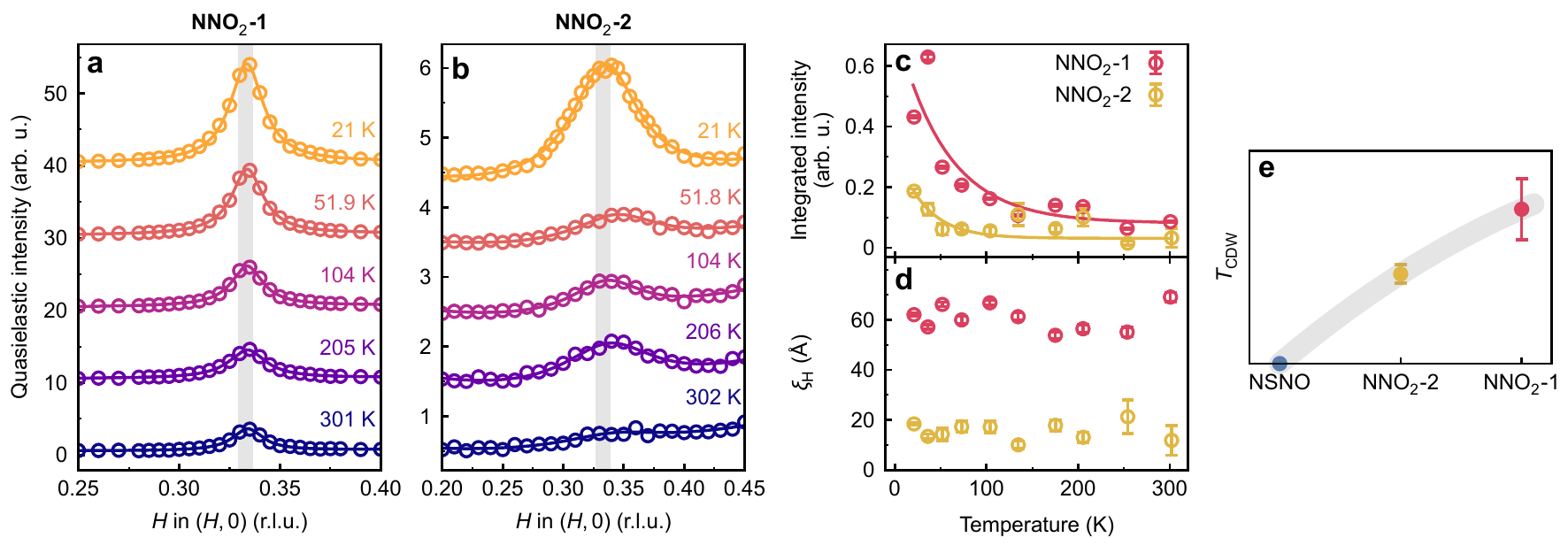}
		\caption{\textbf{Temperature dependence of CDW in NdNiO$_2$.} \textbf{a,} and \textbf{b,} The CDW peaks and their Lorentzian fits along ($H$, 0) direction at various temperatures from NNO$_2$-1, and NNO$_2$-2, respectively. \textbf{c,} Integrated CDW intensity as a function of increased temperature. The temperature dependence is fitted with an exponential function (See Methods). \textbf{d,} CDW in-plane correlation length as a function of increased temperature. \textbf{e,} For a relative comparison, the characteristic CDW temperature is defined as the temperature where the CDW intensity is at $1/e^2$ for NSNO, NNO$_2$-2 and NNO$_2$-1. Error bars in \textbf{c}-\textbf{e} are from least squares fitting.}
		 \label{fig4}
	\end{center}
\end{figure*}

Finally, we performed temperature-dependent RIXS measurements to understand the characteristic temperature of the CDW. Figs.~\ref{fig4}a and b display staggered CDW peaks of NNO$_2$-1 and NNO$_2$-2, respectively. The temperature-dependent CDW intensities decay exponentially from which a critical $T_\mathrm{CDW}$ was obtained (Fig.~\ref{fig4}c), however, no substantial change was seen in the in-plane correlation length (Fig.~\ref{fig4}d). Upon increasing temperature, the CDW intensity decays to a persistent non-zero intensity up to 300\;K, similar to the dynamical CDW fluctuations in cuprates~\cite{miao2017high,arpaia2019dynamical}. It is remarkable that the stronger CDWs retain a relatively high critical temperature $T_\mathrm{CDW}$ as illustrated in Fig.~\ref{fig4}e. The unconventional nature of the CDWs is also implied by already published data, since neither a resistivity anomaly nor a structural phase transition is seen~\cite{li2019superconductivity}. This is in contrast to long-range quasi-static CDWs found in the stripe phase of nickelates, or classical linear-chain compounds~\cite{cheong1994charge,zhang2016stacked, gruner1988dynamics}.


The growth of nickelate superconductors is challenging due to their unstable chemical form, and the susceptibility to structural defects. Thus, it is essential to determine whether the phenomena we see are genuine, rather than a byproduct of growth. The structural analysis of our films reveal a minor contribution of RP defects compared to a vast majority of the infinite-layer phase, despite a small difference in lattice constants (Supplementary Table S1). The O $K$-XAS pre-edge spectra also suggest that the infinite-layer phase dominates the thin films. However, a minor degree of defects and the possible intercalation of hydrogen onto the apical sites may vary in NdNiO$_2$ films, given they went through topotactic reduction under different temperatures. Although the consistent $Q_\mathrm{CDW}$ across three unique NdNiO$_2$ films points out that the defects are unlikely to be the main cause of the symmetry-breaking order, more systematic studies are required to obtain a full picture of the origin. Regarding the case of a structural modulation of the Nd and Ni cations, although there is no obvious hint this is the case from the structural analysis, however it may be induced by a charge modulation via the electron-phonon coupling~\cite{cheong1994charge,comin2016resonant}. It is interesting to note that a previous RIXS study on STO-capped NdNiO$_2$ films did not reveal any CDW signals, but rather magnons, while in the current work on non-capped NdNiO$_2$, we see robust CDWs without sizable magnetic excitations (Supplementary Fig.~S11)~\cite{lu2021magnetic}. These perplexing results suggest STO-capping layers may influence NdNiO$_2$ beyond simple surface protection and calls for systematic investigation.

So far, we have experimentally confirmed the existence of CDWs in infinite-layer NdNiO$_2$ films. More importantly, the involvement of Nd $5d_{xy}$, Nd $5d_{3z^2-r^2}$ and Ni $3d_{x^2-y^2}$ orbitals in the formation of CDWs clearly demonstrates that a minimal multi-orbital model is required to describe the low-energy physics of the infinite-layer nickelates~\cite{lee2004infinite,been2021electronic,sakakibara2020model}. In the cuprates, the low-energy physics are mostly captured by a single hybridised Cu $3d$ and O $2p$ orbital within the CuO$_2$ planes, although it has been shown a multi-orbital model is required~\cite{li2020multiorbital}. Also, unlike CDWs prevalent in the under- or optimally-doped cuprates~\cite{comin2016resonant}, the CDWs are persistent in the parent infinite-layer NdNiO$_2$ films. This, at first sight, may seem incompatible with the half-filling condition, however, the parent compound is slightly self-doped and metallic, effectively mimicking a very underdoped state~\cite{lee2004infinite,hepting2020electronic,been2021electronic}. A broader question is whether CDWs in the infinite-layer nickelates are intertwined with other symmetry-breaking orders, such as AFM order, SDW, or the superconducting state, as in the cuprates~\cite{fradkin2015colloquium}. The missing AFM order may be connected to the competition with the robust CDW state, however the exact cause is yet to be explored~\cite{hayward2003synthesis}. Concerning the relationship between CDWs and superconductivity, our studies are insufficient to make a definitive conclusion, but no CDWs are seen in superconducting Nd$_{0.8}$Sr$_{0.2}$NiO$_2$. Future studies on Sr-doped Nd$_{1-x}$Sr$_{x}$NiO$_2$ with different carrier concentrations are desired. More speculatively, the involvement of Nd $5d_{3z^2-r^2}$ and Ni $3d_{x^2-y^2}$ orbitals close to the Fermi level resembles the situation in some cuprates where the contribution from Cu $3d_{3z^2-r^2}$ orbital works against superconductivity~\cite{sakakibara2010two,peng2017influence}. If the Nd $5d$ states are deemed crucial for the low-energy physics, CDWs may appear more competing than cooperative to superconductivity.  

\textit{Note} - Similar works on the other infinite-layer nickelates appeared at the time of the submission~\cite{Rossi_ABT_2021_, krieger2021charge}

\section*{Acknowledgements}
We thank Mark Dean and Wei-Sheng Lee for insightful discussions. All data were taken at the I21 RIXS beamline of Diamond Light Source (United Kingdom) using the RIXS spectrometer designed, built and owned by Diamond Light Source. We thank Diamond Light Source for providing beam time under proposal ID NT30296. We acknowledge T. Rice for the technical support throughout the experiments. C.C.T. acknowledges funding from Diamond Light Source and the University of Bristol under joint doctoral studentship STU0372. L.Q. and H.L. thank the support from the NSFC (Grant Nos. 11774044, 52072059 and 11822411) and SPRP-B of CAS (Grant No. XDB25000000). K.-J.Z. and H.L. thank the support from NSF of Beijing (Grant No. JQ19002).

\section*{Author contributions}
K.-J.Z. conceived and supervised the project. C.C.T., J.C., K.-J.Z., S.A., M.G.-F., and A.N. performed XAS and RIXS measurements. C.C.T., J.C., and K.-J.Z. analysed RIXS data. L.Q. and X.D. synthesised and characterised thin film samples. M.W. and P.G. performed the STEM measurements. All the authors contributed to the discussion and interpretation of the results. K.-J.Z., C.C.T., and J.C. wrote the manuscript with comments from all the authors.

\section*{Competing interests}
The authors declare no competing interests.

\newpage

\bibliography{References}

\section*{Methods}
\subsection*{Thin film growth}
10 nm thick perovskite NdNiO$_3$ thin films were grown on TiO$_2$-terminated SrTiO$_3$ (001) substrates by pulsed-laser deposition using a 248-nm KrF excimer laser. Prior to the deposition, solid state targets were prepared by sintering and pelletising stochiometric mixtures of NiO and Nd$_2$O$_3$ powder at 1300$^\circ$C for 12 hours. The perovskite Nd$_{0.8}$Sr$_{0.2}$NiO$_3$ thin film is 15 nm thick and was grown using the same laser parameters for the NdNiO$_3$ thin films. Solid state targets were prepared by sintering and pelletising stochiometric mixtures of NiO, Nd$_2$O$_3$ and SrCO$_3$ powder at 1300$^\circ$C for 12 hours~\cite{ding2022stability}. During the deposition, a laser fluence of 1.2 $J/cm^2$ was used to ablate the target and the substrate temperature was controlled at 620$^\circ$C with an oxygen pressure of 200 mTorr. All samples do not have SrTiO$_3$ capping layers. After growth, the thin films were cooled to room temperature under the same oxygen environment. 

Extra care was paid to the growth of perovskite Nd$_{0.8}$Sr$_{0.2}$NiO$_3$ thin films. This was because Sr doping makes the growth of the films more difficult to control than their parent counterpart due to the inclination towards island or layer-island growth rather than epitaxial growth. High quality layer-by-layer growth is necessary to minimise the amount of the defect phases present in order to achieve superconductivity~\cite{lee2020aspects}.

The perovskite samples were reduced to the infinite-layer phase by following a similar topotactic reduction method detailed in Ref.~\cite{li2019superconductivity}. The perovskite films were vacuum-sealed (< 0.1 mTorr) together with 0.1 gram of the solid-state CaH$_2$ powder. The temperature profile of the reduction procedure followed a trapezoidal shape. The warming-up and the cooling-down rates were held at $10^{\circ}$C/min. On the plateau, the reduction was held at a steady temperature for an optimised time of 2 hours. The reduction temperatures of 200$^\circ$C, 220$^\circ$C, and 290$^\circ$C were used to produce the NdNiO$_2$ films denoted NNO$_2$-1, NNO$_2$-3, and NNO$_2$-2, respectively. Nd$_{0.8}$Sr$_{0.2}$NiO$_3$ sample was reduced for 2 hours at an optimised temperature of 300$^\circ$C. NdNiO$_2$ films crystallise tetragonally and are assumed to have in-plane lattice parameters equivalent to SrTiO$_3$  $a = b = 3.91$\;\AA{}. The $c$ lattice parameters were obtained by XRD (Supplementary Note 1).

\subsection*{Thin film characterisation}
Reflection high-energy electron diffraction (RHEED) with a 15 keV electron beam was used to monitor the film quality during growth. The resistivity was measured by a four-probe method via wire-bonded contacts in a cryogen-free magnet system (CFMS, Cryogenic Ltd) down to 1.6 $K$. X-ray reflectivity (XRR), Atomic Force Microscopy (AFM), rocking curve scans and Reciprocal Space Mapping (RSM) of the (002) diffraction peak, and Scanning Transmission Electron Microscopy (STEM) were used to characterise the thin films. XRR, RSM and XRD measurements were performed using a monochromatic source Cu K$\alpha_1$ (Bruker D8 Discover). The surface morphology and roughness of thin films were examined by atomic force microscopy (AFM, NX-10, Park Systems). Atomic-resolution high angle annular dark field scanning transmission electron microscopy (HAADF-STEM) measurements were performed using an aberration-corrected FEI Titan Themis G2 at 300 kV. The STEM specimens were first thinned by mechanical polishing and then subjected to argon ion milling. The ion milling process was carried out using a PIPS (Model 691, Gatan, Inc.). The results of this characterisation are detailed in Supplementary Note 1.

\subsection*{XAS and RIXS measurements}

XAS and RIXS measurements were performed at Beamline I21 at Diamond Light Source, United Kingdom~\cite{zhou2022i21}. The crystallographic $a$–$c$ ($b$–$c$) plane of all the samples were aligned within the horizontal scattering plane (Supplementary Fig.~S8). Reciprocal lattice units (r.l.u.) are defined (where $2\pi/a=2\pi/b=2\pi/c=1$) with $\mathbf{Q}=H\mathbf{a}^{*}+K\textbf{b}^{*}+L\textbf{c}^{*}$. For all the data presented throughout except for scans in $(H,H)$, samples were aligned such that $K = 0$. The polar angular offsets ($\theta$ and $\chi$) of the films were aligned by specular reflection, and the azimuthal offset ($\phi$) by CDW peak, such that the $c^*$ axis was in the scattering plane. The spectrometer arm was at a fixed position of $\Omega=154^\circ$, unless otherwise stated.

XAS spectra were collected with a grazing incidence angle of $\theta_0 = 20^\circ$ to probe both in-plane and out-of-plane unoccupied states. All measurements were done at a temperature of 20\;K with the exit slit opening to 50 $\mu$m at the Ni $L_3$ edge except for the temperature-dependent studies. Fluorescence yield XAS spectra were collected with a photodiode and normalised to the incoming beam intensity. Both linear vertical ($\sigma$) and horizontal ($\pi$) polarisations were used. While $\sigma$ polarised light purely probes the in-plane XAS, \textit{i.e.}, $I_\text{ab} = I_\sigma$, the out-of-plane XAS were obtained by a combination of both $\sigma$ and $\pi$ polarisations with $I_\text{c} = \left( I_\pi - I_\sigma\sin^2[\theta_0] \right) / \cos^2[\theta_0]$.

Energy-dependent RIXS measurements were performed at an in-plane position of $Q=(-0.35,0)$ at a temperature of 20\;K with the exit slit opening to 30 $\mu$m corresponding to an average energy resolution of 41 meV (FWHM). The incident energy range went from 851.5 to 854\;eV in steps of 100\;meV to fully capture the resonance behaviour across the Ni-$L_3$ absorption peaks.

Momentum-dependent RIXS measurements were performed at resonant energies determined by the absorption peaks in XAS at a temperature of 20\;K with the exit slit opening to 40 $\mu$m corresponding to an average energy resolution of 53 meV (FWHM). To maximise the CDW signal, we used $\sigma$ polarisation and a grazing out geometry ($\theta > \Omega/2$). $L$ dependent RIXS measurements were done by positioning the spectrometer arm to different $\Omega$ angles such that the CDW was always centered at $H = 0.333$ $r.l.u$.

\subsection*{Data fitting}

RIXS data were normalised to the incident photon flux, and subsequently corrected for self-absorption effects prior to fitting. Energy calibration was obtained by fitting the quasielastic peak to a pseudovoigt with width fixed to instrument resolution and setting the centre of the peak to $E=0$. The quasielastic peak intensities presented throughout the text were taken by integrating the RIXS spectra in the range $\left[-27, 27 \right]\;$meV.

$H$ dependence of the quasielastic peak intensity was fitted to a Lorentzian peakshape atop a linear background. The in-plane correlation length is defined as $\xi_H=1/\Gamma$ at the condition of $\Omega=154^\circ$, where $\Gamma=\text{HWHM}$ is the scale parameter of the fitted Lorentzian. Note that $\xi_H$ may not be equal but should be comparable to the absolutely genuine correlation length of the CDW object as its entire 3D shape is unknown. To convert from \AA{}$^{-1}$ to r.l.u., we used the SrTiO$_3$ lattice parameter $a=3.91\;$\AA{}. Temperature dependence of the integrated CDW intensity was fitted to an exponential function of the form $I(T;a,b,c) = a \exp(-bT) + c$. The $1/e^2$ intensity was $2/b$.

XAS data were normalised to the incident photon flux and then the pre- $L_3$-edge intensity was aligned to zero with a fixed intensity at the post-edge for all samples. XAS were collected up to the $L_2$-edge and the $I_\text{c}$ projection was normalised to the $I_\text{ab}$ by scaling the $L_3$ post-edge intensity by a constant factor minimising the difference in intensities by least squares.

The features of the $L_3$ edge peak and post edge were fitted with four Gaussian peaks and an arctan step function. The centres of the peaks were fixed at approximately 852.2, 852.8, 853.7, and 856.3\;eV and the step was fixed at 860.7\;eV. Small changes were allowed in the peak centre between samples to improve the reduced $\chi^2$ but were fixed between polarisations of the same sample. From left to right of each XAS spectrum, the width of the first two Gaussians were fixed to each other, and the width of the final Gaussian and step were fixed to each other. All other parameters were left floating. The peaks of interest were centred around $\sim 852.2\;$eV and $852.8\;$eV. We assigned the lower energy peak to one from Nd-Ni hybridisation and higher energy peak to the Ni$^{1+}$ $2p\rightarrow3d$ transition. The Nd 5$d$ and Ni 3$d$ orbital contents were calculated by taking the ratio fitted peak intensities of the out-of-plane XAS and the total fitted peak intensity of both projections, \textit{i.e.} $\text{orbital content} = I_\text{c}/\left(I_\text{c} + I_\text{ab}\right)$. This was done for the Nd-Ni hybridised peak and Ni-$L_3$ peak separately to determine the Nd $5d_{3z^2-r^2}$ and Ni $3d_{3z^2-r^2}$ orbital content respectively.

\section*{Data availability}
All the data supporting the findings of this study are available in the Supplementary Information and deposited in the Zenodo repository at \url{https://doi.org/10.5281/zenodo.6778273}. Further information is available from the corresponding authors on reasonable request.


\end{document}


\title{Supplementary information for ``Charge density waves in infinite-layer NdNiO$_2$ Nickelates''}

\author{Charles C. Tam}
    \thanks{These authors contributed equally}
    \affiliation{Diamond Light Source, Harwell Campus, Didcot OX11 0DE, United Kingdom}
    \affiliation{H. H. Wills Physics Laboratory, University of Bristol, Bristol BS8 1TL, United Kingdom}

\author{Jaewon Choi}
    \thanks{These authors contributed equally}
    \affiliation{Diamond Light Source, Harwell Campus, Didcot OX11 0DE, United Kingdom}
    
\author{Xiang Ding}
    \thanks{These authors contributed equally}
    \affiliation{School of Physics, University of Electronic Science and Technology of China, Chengdu, 610054, China}

\author{Stefano Agrestini}
    \affiliation{Diamond Light Source, Harwell Campus, Didcot OX11 0DE, United Kingdom}

\author{Abhishek Nag}
    \affiliation{Diamond Light Source, Harwell Campus, Didcot OX11 0DE, United Kingdom}
    \affiliation{Laboratory for Non-linear Optics, Paul Scherrer Institut, CH-5232 Villigen, Switzerland}
    
\author{Mei Wu}
    \affiliation{International Center for Quantum Materials and Electron Microscopy Laboratory, School of Physics, Peking University, Beijing 100871, China}

\author{Bing Huang}
    \affiliation{Beijing Computational Science Research Center, Beijing 100193, China}

\author{Huiqian Luo}
    \affiliation{Beijing National Laboratory for Condensed Matter Physics, Institute of Physics, Chinese Academy of Sciences, Beijing 100190}
    \affiliation{Songshan Lake Materials Laboratory, Dongguan, Guangdong 523808, China}
    
\author{Peng Gao}
    \affiliation{International Center for Quantum Materials and Electron Microscopy Laboratory, School of Physics, Peking University, Beijing 100871, China}
\author{Mirian Garc\'ia-Fern\'andez}
    \affiliation{Diamond Light Source, Harwell Campus, Didcot OX11 0DE, United Kingdom}

\author{Liang Qiao}
    \email{liang.qiao@uestc.edu.cn}
    \affiliation{School of Physics, University of Electronic Science and Technology of China, Chengdu, 610054, China}

\author{Ke-Jin Zhou}
    \email{kejin.zhou@diamond.ac.uk}
    \affiliation{Diamond Light Source, Harwell Campus, Didcot OX11 0DE, United Kingdom}


\maketitle

\thispagestyle{empty}

\newpage

\section*{Supplementary Note 1: Thin film characterisation} \label{sec:characterisation}

In this section we present physical properties of three parent NdNiO$_2$ films (NNO$_2$-1, NNO$_2$-2, and NNO$_2$-3), and the superconducting Nd$_{0.8}$Sr$_{0.2}$NiO$_2$ film (NSNO) that we studied. These films were prepared under slightly different conditions (see the Method for details).

\begin{figure}[!htb]
	\centering
		\includegraphics[width=\textwidth]{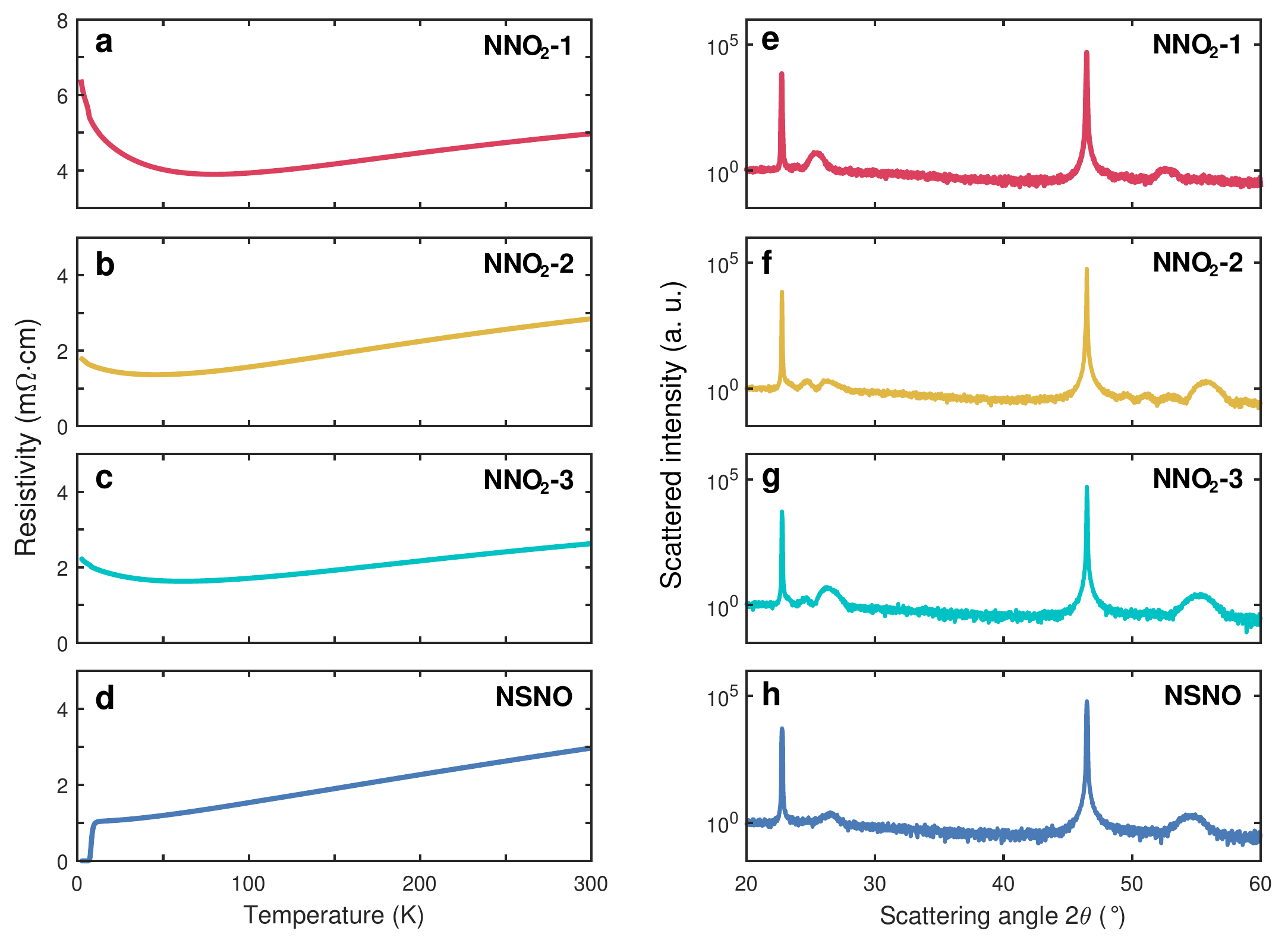}
		\caption{\textbf{Resistivity and XRD data of the parent NdNiO$_2$ and the superconducting Nd$_{0.8}$Sr$_{0.2}$NiO$_2$ (NSNO) thin films.} Note that the parent NdNiO$_2$ thin films show weakly insulating behaviour and the doped NSNO has a superconducting onset temperature of about 10$\;K$. }
		\label{Ext_Fig1}
\end{figure}

Fig.~\ref{Ext_Fig1} shows the resistivity and the X-ray diffraction data of all the samples. The NdNiO$_2$ films show a characteristic weakly diverging resistivity at low temperature, which has been previously reported in the infinite-layer nickelates (e.g. Ref.~\cite{li2019superconductivity}). The $c$ lattice parameters of the samples are extracted from XRD and listed in Table~\ref{table:clatt}. 

\begin{figure}[!htb]
	\begin{center}
		\includegraphics[width=.8\textwidth]{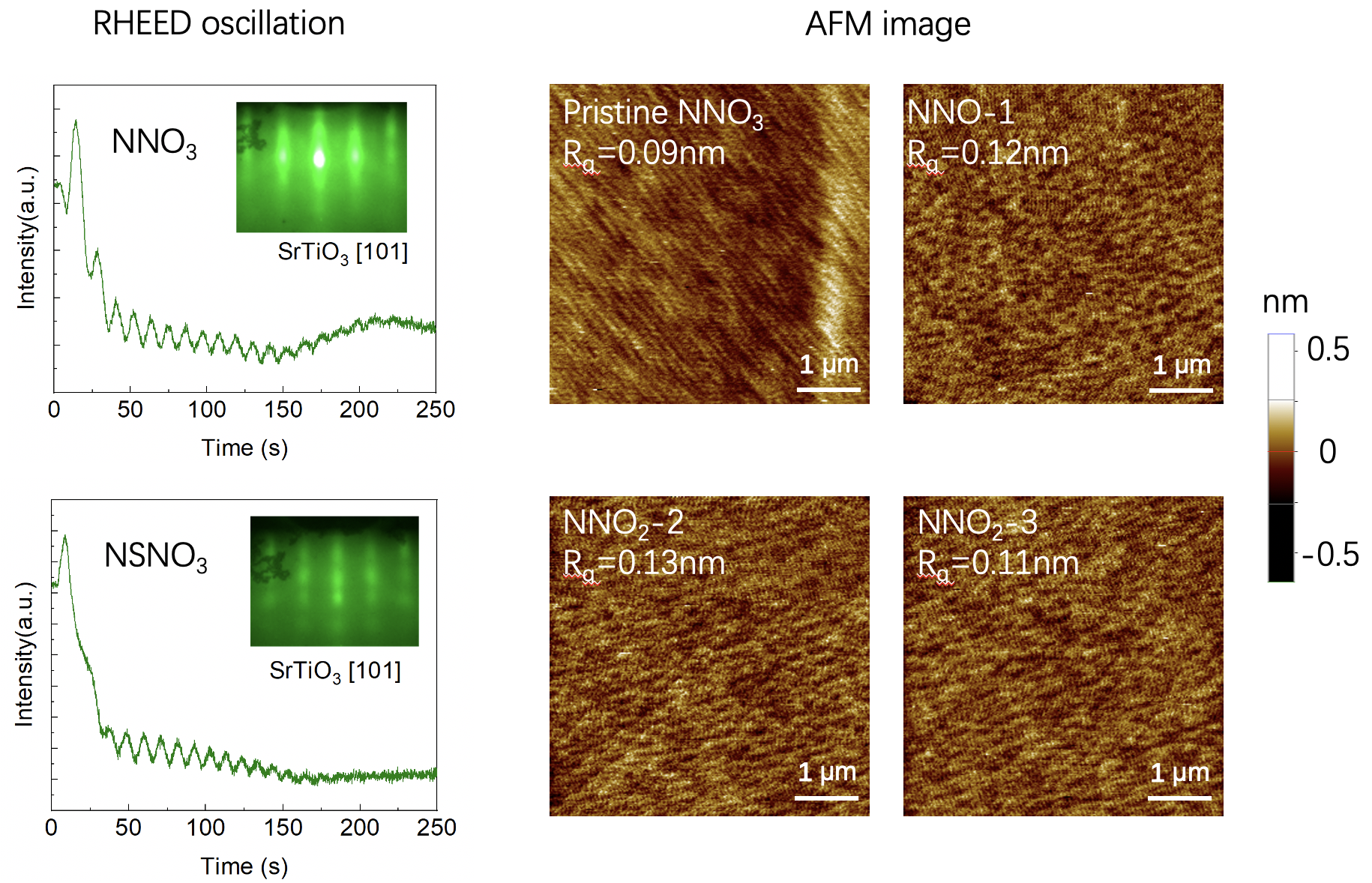}
		\caption{\textbf{RHEED patterns of the pristine NdNiO$_3$ (NNO$_3$) and Nd$_{1-x}$Sr$_{x}$NiO$_3$ (NSNO$_3$) samples. AFM images of the pristine NNO$_3$ and all NNO$_2$ samples.} The RHEED patterns demonstrate that NdNiO$_3$ and Nd$_{0.8}$Sr$_{0.2}$NiO$_3$ films were grown in a layer-by-layer mode. The AFM images show the root mean square of the average height deviation (R$_q$) in the order of 0.1 nm demonstrating a good surface quality for all samples.}
		 \label{fig:RHEED_AFM}
	\end{center}
\end{figure}

\begin{figure}[!htb]
	\begin{center}
		\includegraphics[width=.8\textwidth]{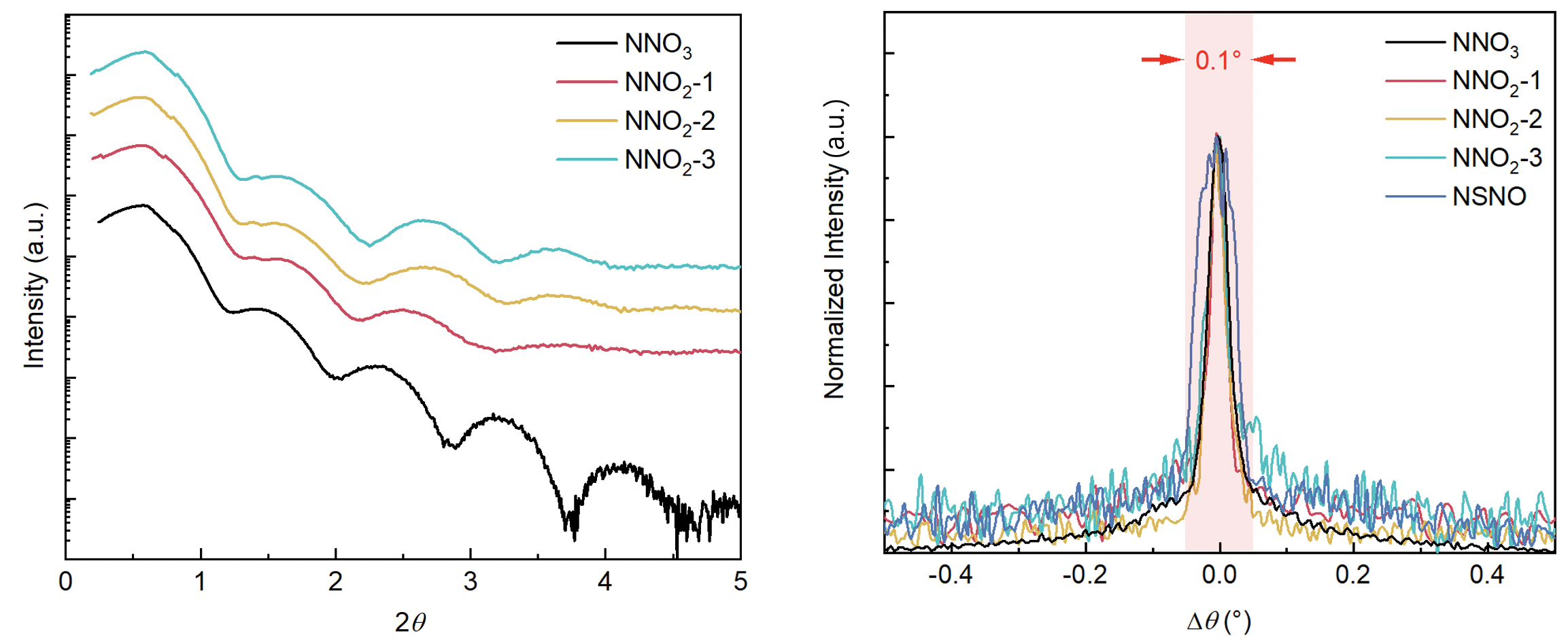}
		\caption{\textbf{X-ray reflectivity and rocking curve scans across the (002) diffraction peak from the pristine NNO$_3$, NNO$_2$-1, NNO$_2$-2, NNO$_2$-3, and NSNO samples.} X-ray reflectivity data show a clear change of the oscillation period from NdNiO$_3$ to NdNiO$_2$ due to the change of the crystal structure following the topotactic reduction. The rocking curve scans across the (002) diffraction peak demonstrate that the crystal structure of the films are similar.}
		 \label{fig:XRR_RC}
	\end{center}
\end{figure}

\begin{figure}[!htb]
	\begin{center}
		\includegraphics[width=0.9\textwidth]{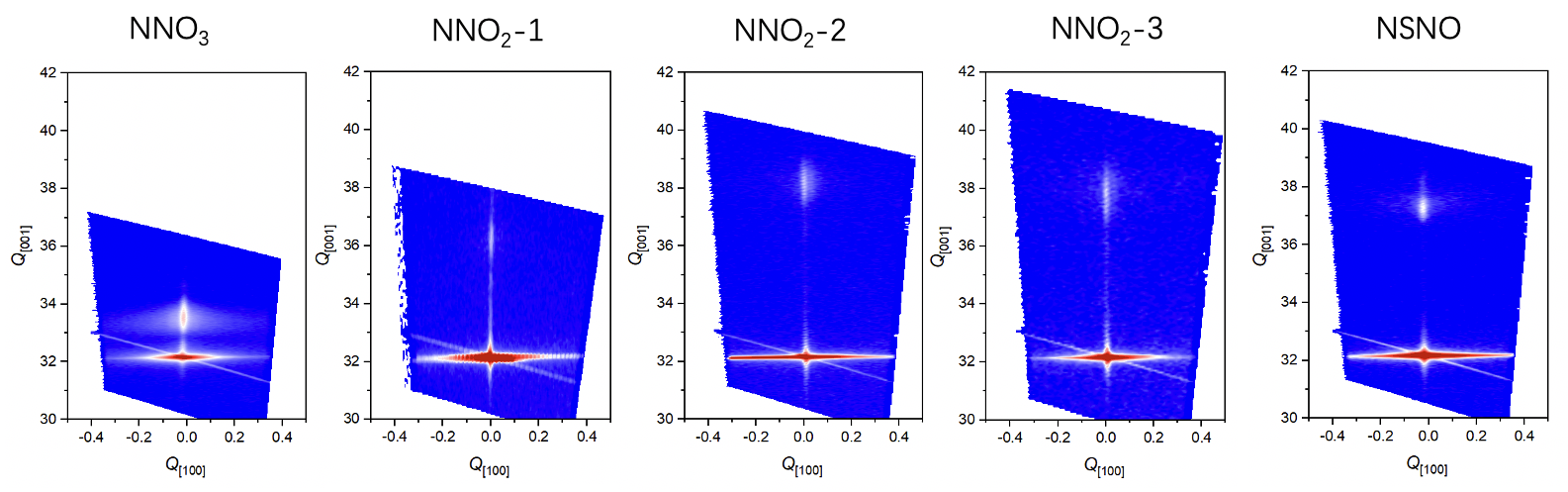}
		\caption{\textbf{Reciprocal Space Mapping of the (002) diffraction peak from the pristine NNO$_3$, NNO$_2$-1, NNO$_2$-2, NNO$_2$-3, and NSNO samples.} The RSM results show that the (002) diffraction peak has similar width suggesting no drastic differences in terms of the film quality. Note that the RSM data yield the same c-axis lattice constants as retrieved from the XRD data.}
		 \label{fig:RSM}
	\end{center}
\end{figure}

The X-ray reflectivity (XRR), reciprocal space mapping (RSM) and rocking curve scans of the (002) diffraction peak provide information about the thin film samples in reciprocal space. Reflected high-energy electron diffraction (RHEED) patterns demonstrate that NdNiO$_3$ and Nd$_{1-x}$Sr$_{x}$NiO$_3$ films were grown in a layer-by-layer mode (Fig.~\ref{fig:RHEED_AFM}). XRR data show a clear change of the oscillation period from NdNiO$_3$ to NdNiO$_2$ due to the change of the crystal structure following the topotactic reduction procedure (Fig.~\ref{fig:XRR_RC}). The rocking curve scans across the (002) diffraction peak demonstrate that all films have similar crystalline structure (Fig.~\ref{fig:XRR_RC}). The RSM results show that the (002) diffraction peak has a comparable width across all the samples suggesting again no drastic differences in terms of the film quality (Fig.~\ref{fig:RSM}). Note that the RSM data yield the same c-axis lattice constants as retrieved from the XRD data.     

\begin{figure}[!htb]
	\begin{center}
		\includegraphics[width=0.8\textwidth]{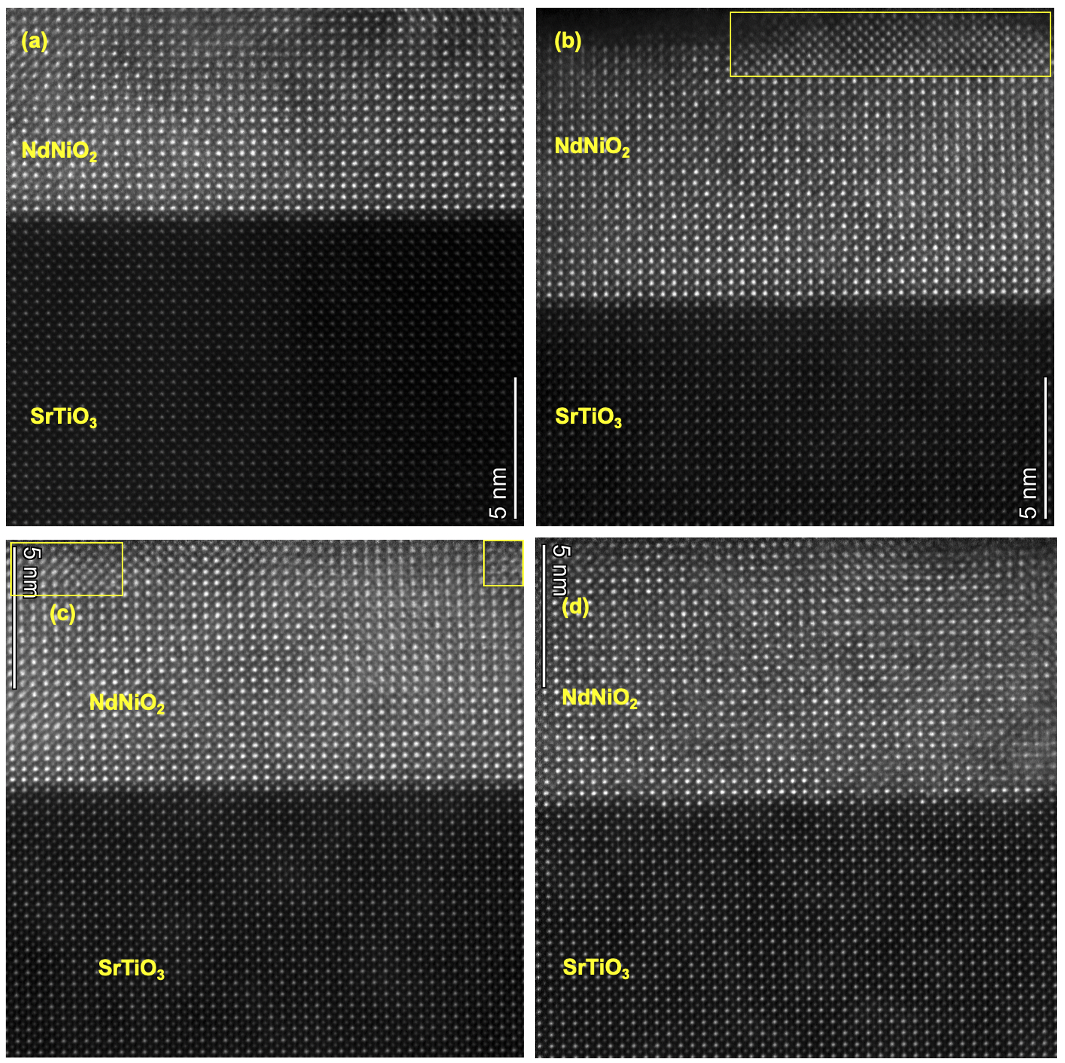}
		\caption{\textbf{STEM images of NNO$_2$-1 sample (a, b) and NNO$_2$-2 sam ple (c, d).} The yellow boxes highlight the minor RP defect phase while the majority of the probed region shows the square-planar 112 phase.}
		 \label{fig:TEM_S11S15_part}
	\end{center}
\end{figure}

From real space AFM images, we obtained the root mean square of the average height deviation (R$_q$) in the order of 0.1 nm demonstrating a good surface quality for all samples (Fig.~\ref{fig:RHEED_AFM}). Analysing the STEM images reveals that the majority of NNO$_2$-1 and NNO$_2$-2 are made of the infinite square-planar layered structure (Fig.~\ref{fig:TEM_S11S15_part}). However, there are some small regions where the Ruddlesden-Popper (RP) secondary phase is present (highlighted by yellow boxes). Areas of RP defect tend to form close to the surface of the film~\cite{lee2020aspects}. 

The c-axis lattice constant of the NNO$_2$-1 sample is larger than that of NNO$_2$-2 and NNO$_2$-3 despite having comparable crystalline quality and vast majority of 112 phase present inferred from the RSM, rocking curves, the STEM results, and the O $K$-XAS spectra. The exact cause is unclear however our observation suggests the 113 to 112 structural transition might not be as simple as previously thought, namely the perovskite ABO$_3$ structure can be directly converted into the infinite-layer ABO$_2$ structure. The end product of the reduction depends on multiple parameters, such as the quantity of CaH$_2$, the reduction temperature, the reduction time, and the potential intercalation of hydrogen. We suspect that the transition from 113 to 112 phase might involve some intermediate process or a possible structural transition which deserves a further in-depth study.

\FloatBarrier

\begin{table}[!htb]
\centering
\begin{tabular}{@{}cc@{}}
\toprule
Sample name & $c$ lattice parameter (\AA) \\ \midrule
NNO$_2$-1 & 3.475 \\
NNO$_2$-2 &  3.295  \\
NNO$_2$-3 & 3.326 \\
NSNO & 3.363  \\ \bottomrule
\end{tabular}
\caption{$c$ lattice parameters extracted from XRD.}
\label{table:clatt}
\end{table}

\FloatBarrier

\FloatBarrier

\section*{Supplementary Note 2: Estimation of the inhomogeneity via O $K$-XAS spectra} \label{sec:homogeneity}

In this section we present an estimation of the sample inhomogeneity by and analysis of the O $K$-edge X-ray absorption spectroscopy (XAS) spectra. As demonstrated by electron energy loss spectroscopy and XAS, the pre-peak of the O $K$-edge is very sensitive to the local electronic structure of the perovskite and square-planar nickelates~\cite{goodge2021doping,pan2022superconductivity,lin2021strong}. 

\begin{figure}[!htb]
	\begin{center}
		\includegraphics[width=0.9\textwidth]{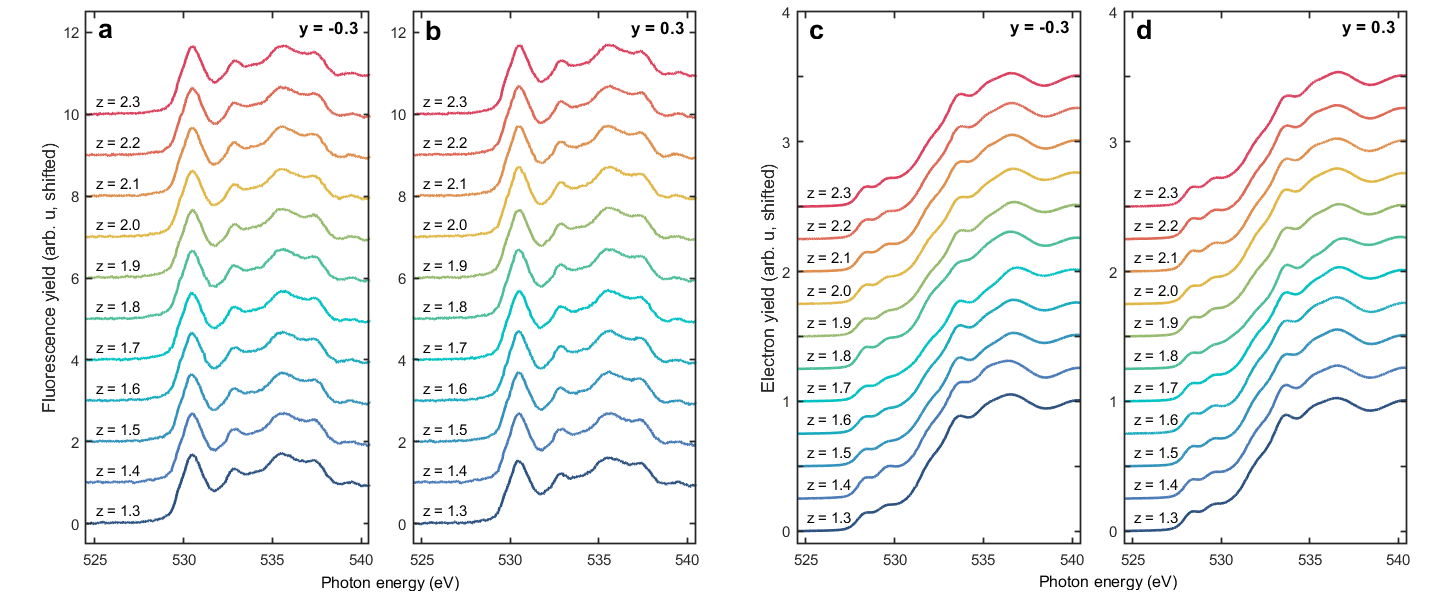}
		\caption{\textbf{The O $K$-XAS spectra for the inhomogeneity study.} XAS spectra were measured in a mesh across a region of 1.5 mm by 1.0 mm around the centre of the NSNO sample. The grazing incident angle is 20$^\circ$ relative to the sample surface. y (z) represents the length (width) of the mesh. All other parameters of the experimental setup is the same as those for the NdNiO$_2$ samples.}
		 \label{fig:homogeneity_NSNO}
	\end{center}
\end{figure}

\begin{figure}[!htb]
	\begin{center}
		\includegraphics[width=0.7\textwidth]{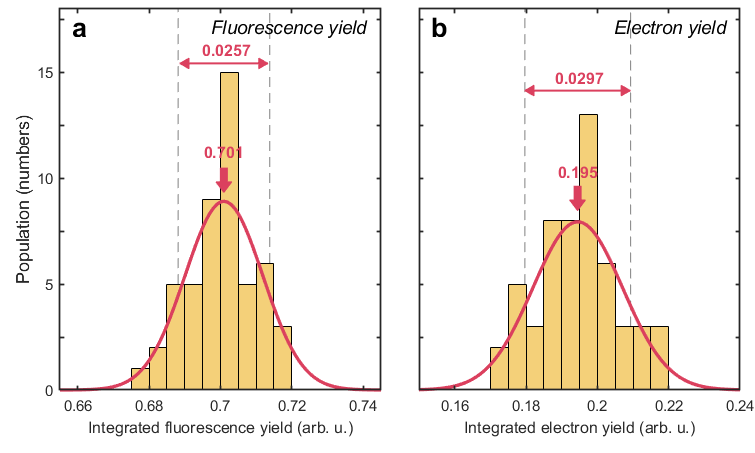}
		\caption{\textbf{Histograms of the inhomogeneity of the NSNO sample.} Histograms are extracted from the bulk-sensitive Fluorescence yield (a) and the surface-sensitive electron yield (b) O $K$-XAS spectra.}
		 \label{fig:homogeneity_NSNO_histogram}
	\end{center}
\end{figure}

Fig.~\ref{fig:homogeneity_NSNO} summarises some representative O $K$- XAS spectra from both the bulk-sensitive fluorescence yield and the surface-sensitive electron yield mesh measurements conducted in a region of 1.5 mm by 1.0 mm around the centre of the sample. The sample has physical dimensions of 2.0 mm $\times$ 2.0 mm. Overall the spectra look highly comparable. The pre-peak regime (527-530 eV) was integrated for all 50 spectra. Histograms of the the spectral weight of the pre-peak are shown in Fig.~\ref{fig:homogeneity_NSNO_histogram}. The full width at half maximum (2.35$\sigma$) extracted from the analysis is 3.7\% and 15.2\% relative to the mean value of the bulk and the surface, respectively, as a representation of the level of the inhomogeneity. Interestingly, the inhomogeneity of the surface is worse than the bulk in accord with the experience that the surface is more likely to form RP defect phase than the bulk. We estimate that the NdNiO$_2$ thin film samples have similar levels of inhomogeneity as the Nd$_{0.8}$Sr$_{0.2}$NiO$_2$ film because of their comparable O $K$-XAS spectra (Fig.~\ref{fig:ok-xas}). Note that the NdNiO$_2$ films were reduced at quite different temperatures however show very similar pre-edge intensity in the bulk and at the surface. This suggests that the reduction from 113 to 112 phase is nearly completed for all NdNiO$_2$ samples. 

\newpage

\section*{Supplementary Note 3: Additional RIXS}

\begin{figure}[h]
	\begin{center}
		\includegraphics[width=.5\textwidth]{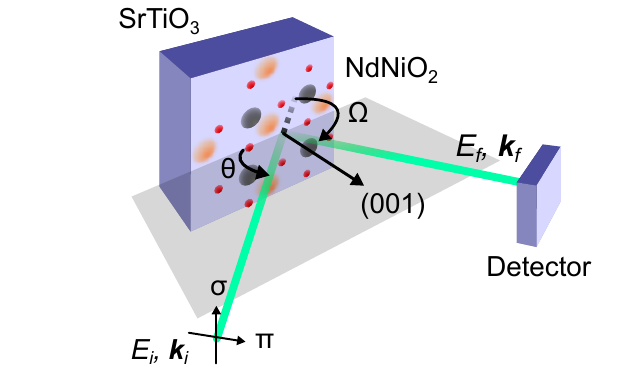}
		\caption{\textbf{The experimental geometry used for the XAS and RIXS measurements.} $E_i$ ($E_f$) and $\textbf{k}_i$ ($\textbf{k}_f$) represents the incoming (outgoing) photon energy and momentum, respectively. $\sigma$ and $\pi$ stands for the incoming linear vertical and linear horizontal polarisation, respectively. $\theta$ is the grazing incident angle between the sample surface and the incoming x-rays while $\Omega$ represents the scattering angle of RIXS experiments.}
		 \label{Ext_Fig2}
	\end{center}
\end{figure}

We present here additional RIXS measurements to support the main text. The scattering geometry employed for XAS and RIXS measurements is illustrated in Fig.~\ref{Ext_Fig2}. 

\begin{figure}[h]
	\begin{center}
		\includegraphics[width=\textwidth]{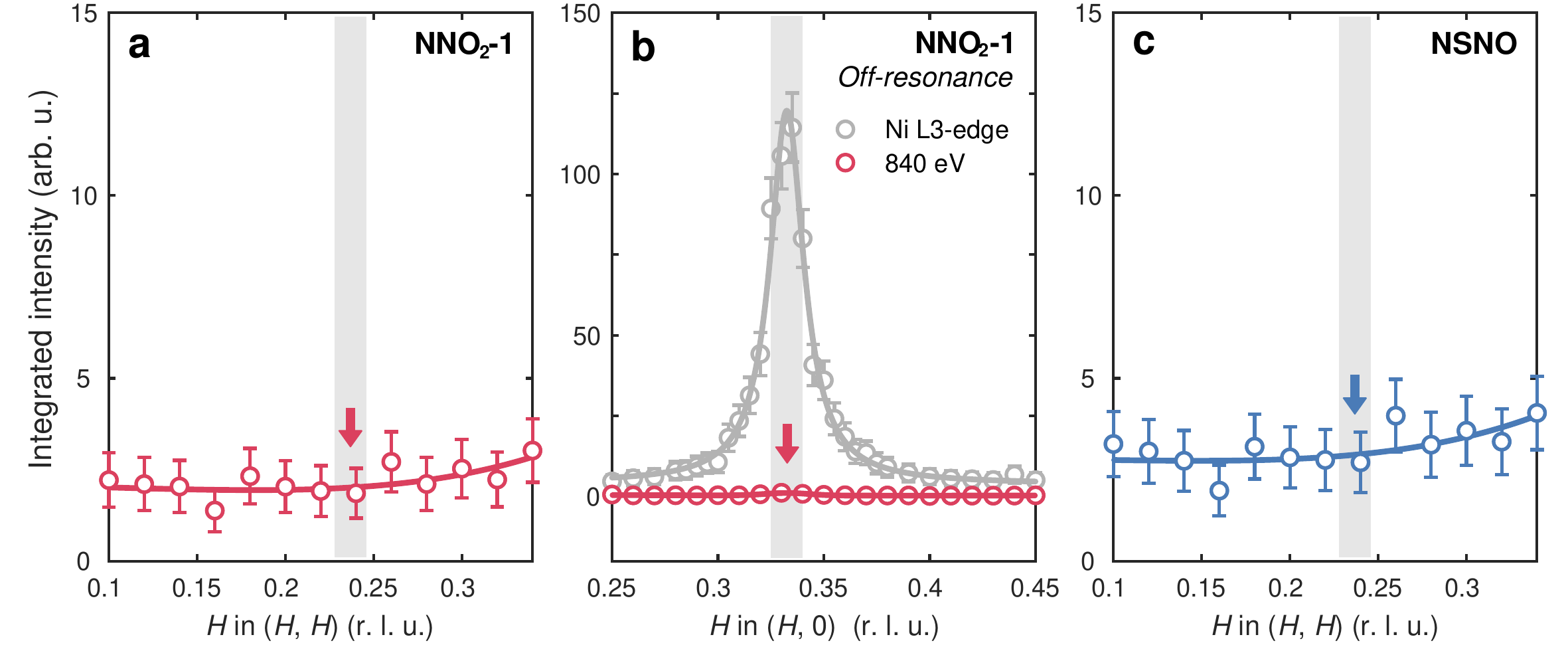}
		\caption{\textbf{The CDWs behaviour in NNO$_2$-1 and NSNO.} \textbf{a} The absence of the CDW peak along ($H$, $H$) direction in NNO$_2$-1. \textbf{b,} The CDW peak in NNO$_2$-1 along ($H$, 0) direction at the on- and off- Ni $L_3$ resonance. \textbf{c,} The absence of the CDW peak along ($H$, $H$) direction in NSNO.}
		 \label{Ext_Fig4}
	\end{center}
\end{figure}

Additional characterisation of the CDW of NNO$_2$-1 and the lack of CDW in NSNO are plotted in Fig.~\ref{Ext_Fig4}. Scans along $(H,H)$ (45$^\circ$ to the Ni-O bond direction) of NNO$_2$-1 (Fig.~\ref{Ext_Fig4}a) and NSNO (Fig.~\ref{Ext_Fig4}c) show no sign of CDW, indicating CDW modulations are mostly parallel to the Ni-O bonds. Fig.~\ref{Ext_Fig4}b shows a scan of NNO$_2$-1 $\sim10$ eV below the resonance, indicating what we see in the main text is indeed specific to Ni $3d$ states and the Nd-Ni hybridised states.

\begin{figure}[h]
	\begin{center}
		\includegraphics[width=\textwidth]{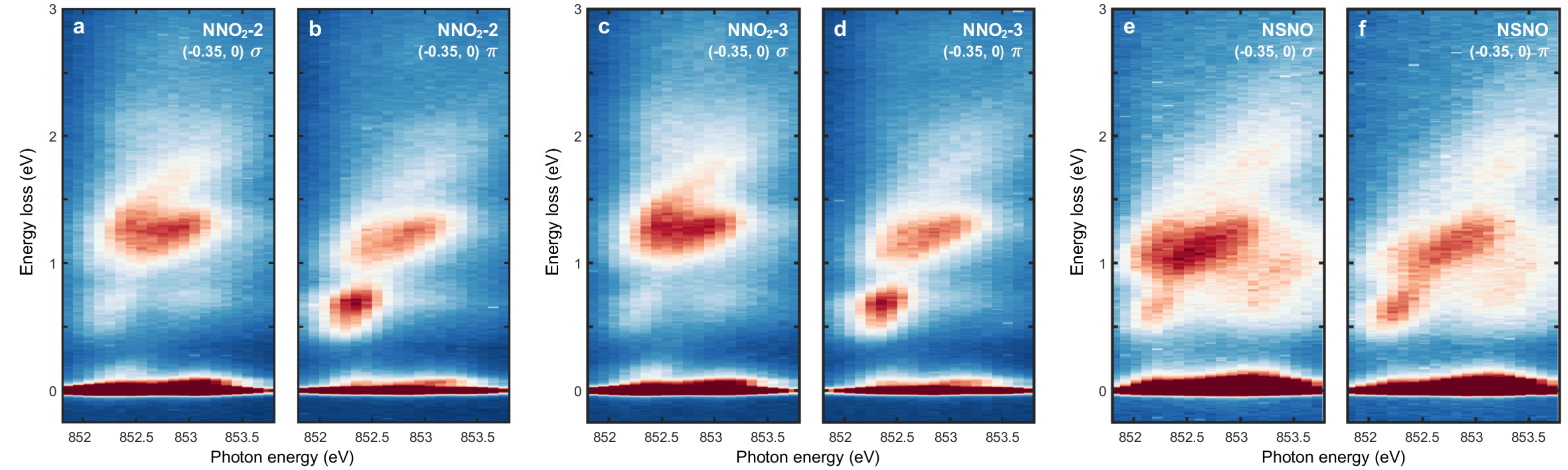}
		\caption{\textbf{RIXS energy maps of NNO$_2$-2, NNO$_2$-3 and NSNO.} Note that all RIXS maps were collected at (-0.35, 0) in both polarisations.}
		 \label{Ext_Fig6}
	\end{center}
\end{figure}

Energy detuned RIXS of the sample NNO$_2$-1 is presented in Fig.1 of the main text. Fig.~\ref{Ext_Fig6} shows all the energy detuned RIXS measured at both $\pi$ and $\sigma$ polarisations for the other three samples.

\begin{figure*}[!ht]
	\begin{center}
		\includegraphics[width=.9\textwidth]{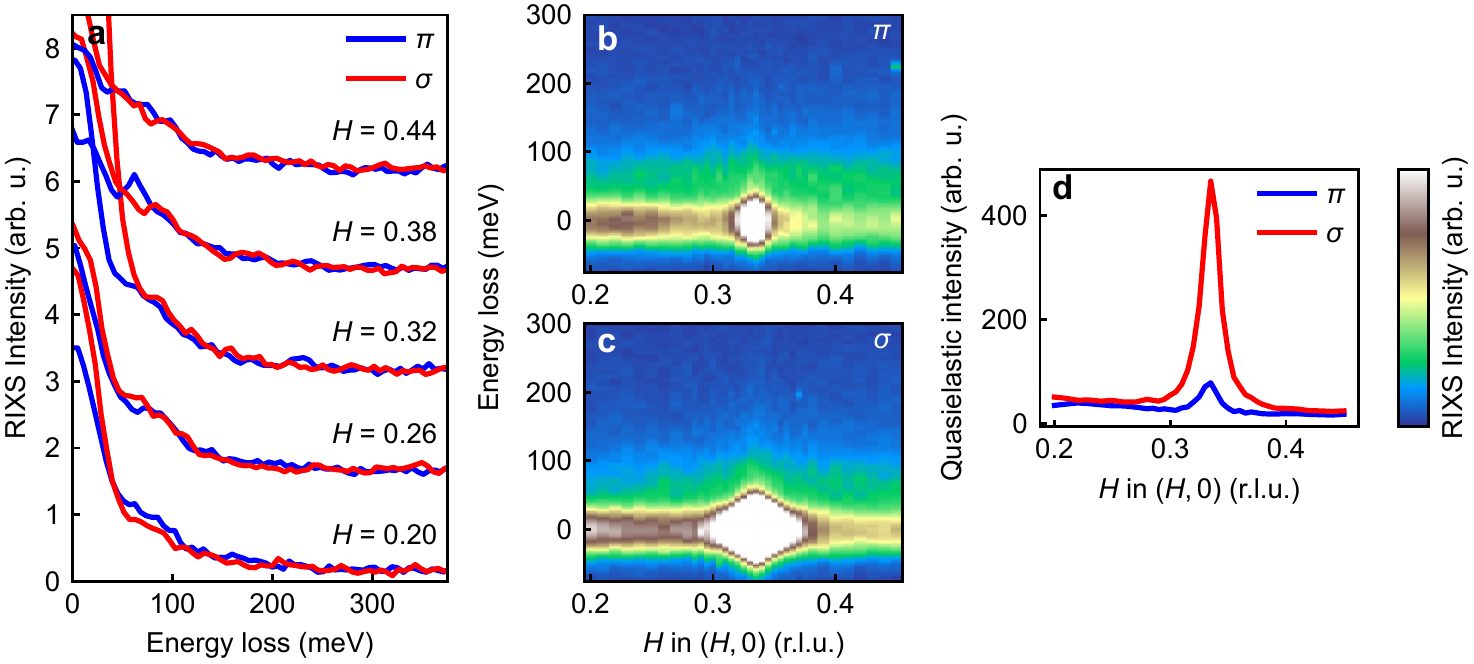}
		\caption{\textbf{Momentum-dependent RIXS of NdNiO$_2$.} RIXS spectra at various in-plane momentum transfer values of NNO$_2$-1 in (\textbf{a}) measured with both $\pi$ and $\sigma$ polarised incident x-rays. (\textbf{b})-(\textbf{c}) RIXS colourmaps of the same measurements. (\textbf{d}) Quasielatic intensity, integrated in $[-27,27]\;$meV and normalised to $dd$ excitations in $[0.9,1.5]\;$eV, of the two polarisations.}
		 \label{fig:polarisation}
	\end{center}
\end{figure*}
\begin{figure*}[!ht]
	\begin{center}
		\includegraphics[width=.5\textwidth]{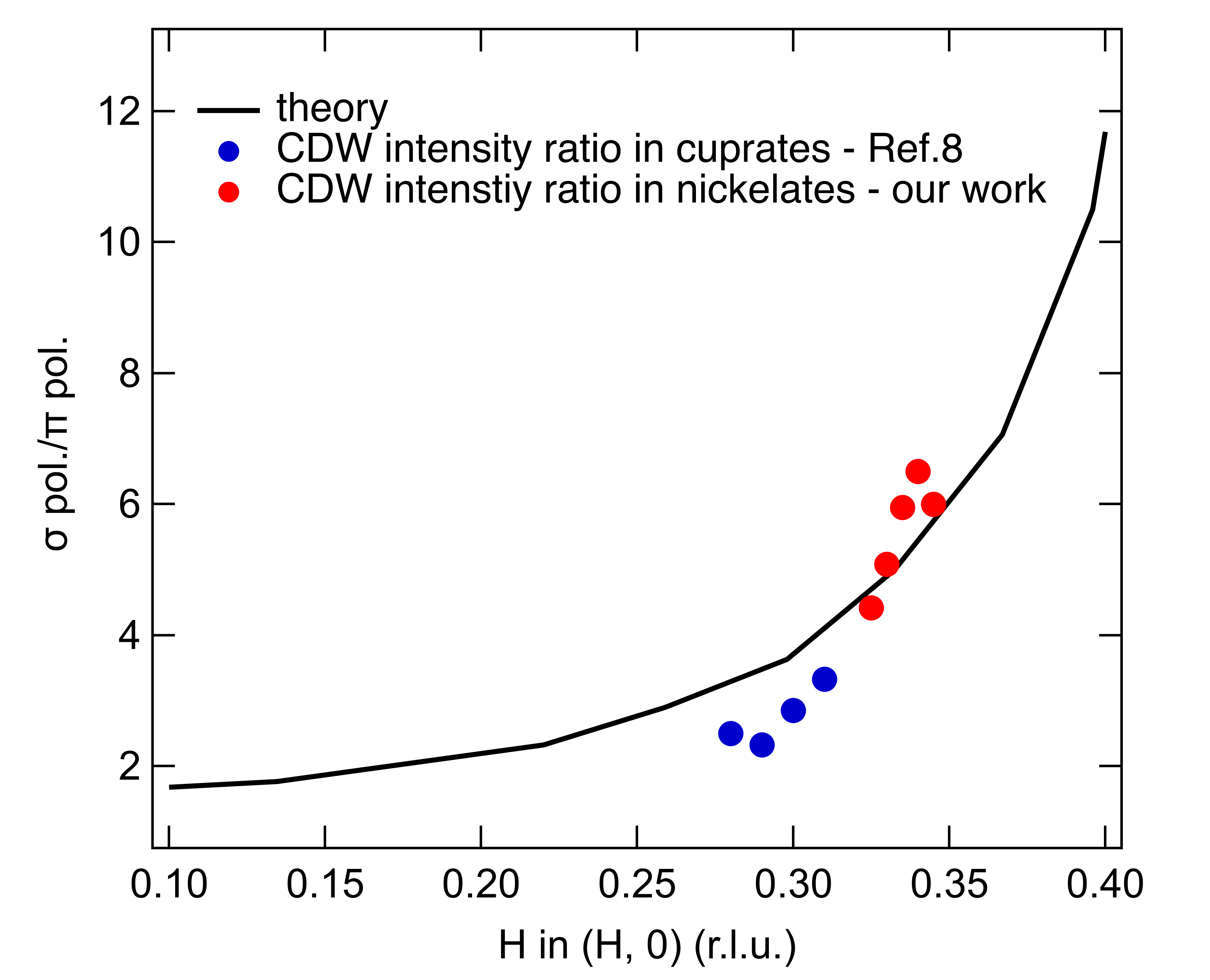}
		\caption{\textbf{CDW polarisation intensity ratio.} $\sigma/\pi$ CDW polarisation ratio of an infinite-layer nickelate NdNiO$_2$ from this work in red, theory in black, and YBCO in blue, from Ref.~\cite{ghiringhelli2012long}.}
		 \label{fig:cdw_ratio}
	\end{center}
\end{figure*}

Fig.~\ref{fig:polarisation}a shows momentum-dependent RIXS along $(H,0)$ of NNO$_2$-1 demonstrating the low-energy excitations at both $\sigma$ and $\pi$ polarisations. All the RIXS spectra describe essentially no evidence of sizeable inelastic excitations in the region $100 < E < 500$\;meV. The low-energy excitation is in stark contrast to the SrTiO$_3$-capped NdNiO$_2$ thin film where a dispersive magnon is observed~\cite{lu2021magnetic}. Fig.~\ref{fig:polarisation}b and c show the momentum-dependent RIXS maps in which a quasielastic scattering peak is centered at a wavevector of $\sim(1/3, 0)$.  In Fig.~\ref{fig:polarisation}d is plotted the integrated quasielastic intensity in an energy window of [-27, 27] meV. The peak has an enhancement in $\sigma$ polarisation than $\pi$ polarisation. In Fig.~\ref{fig:cdw_ratio} we plot the calculated ratio of $\sigma$ to $\pi$ polarised charge scattering based on a single-ion $3d_{x^2-y^2}$ model~\cite{ament2009theoretical}. We also superimpose on theoretical curve the experimental ratio of the electronic Bragg peak near the Q$_{CDW}$ wavevector of the infinite-layer nickelates as well as that of CDW in the cuprate YBa$_2$Cu$_3$O$_{6+\delta}$~\cite{ghiringhelli2012long}. Here we see the polarisation ratio of the CDW in both the nickelate and the cuprate agree within error with theory.

\FloatBarrier

\section*{Supplementary Note 4: Additional XAS}

\begin{figure}[h]
	\begin{center}
		\includegraphics[width=\textwidth]{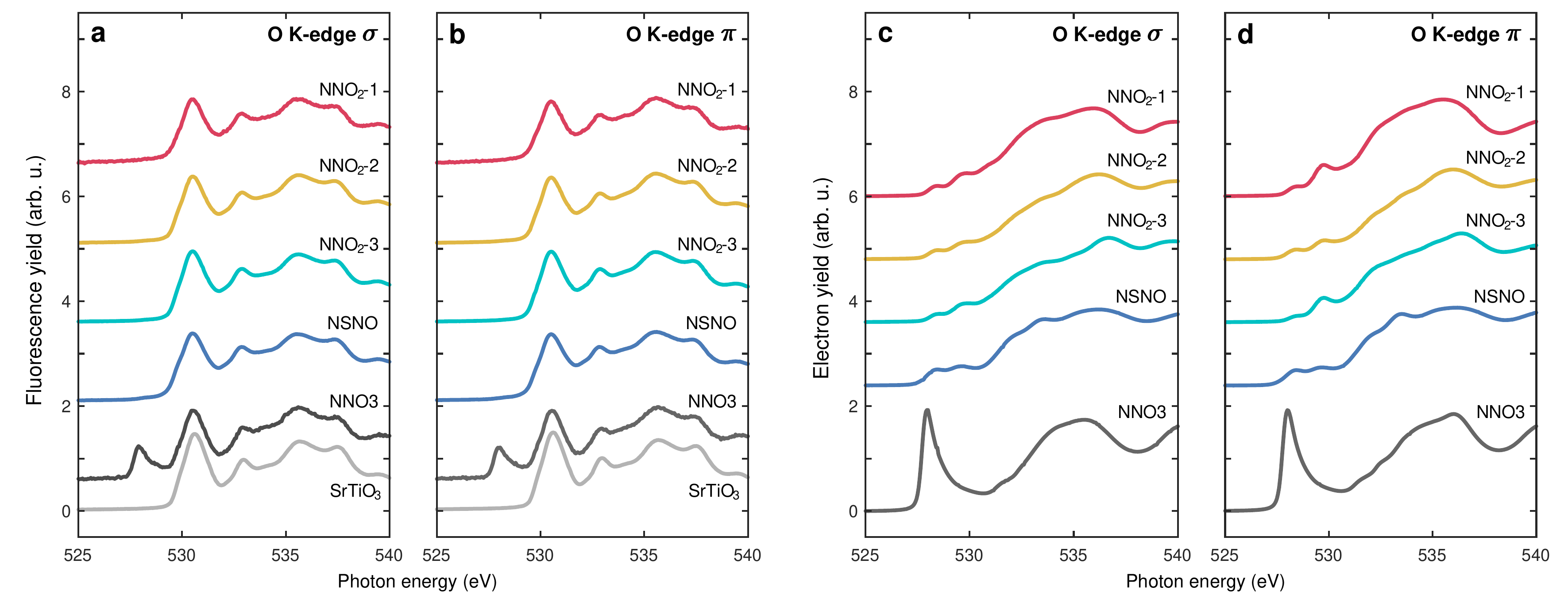}
		\caption{\textbf{O $K$ XAS spectra of thin film samples as well as as well as the as-grown NdNiO$_3$ thin film.} \textbf{a (b),} O $K$ XAS spectra in $\sigma$ ($\pi$) incoming polarisation collected using fluorescence yield representing the signal from the bulk. \textbf{c (d),} O $K$ XAS spectra in $\sigma$ ($\pi$) incoming polarisation collected using total electron yield representing the signal from the surface layers. In addition to the main samples, we also present O $K$ XAS data from the reference samples NdNiO$_3$ (NNO3) for both the electron yield and the fluorescence yield and the O $K$ XAS data from the SrTiO$_3$ single crystal.}
		 \label{fig:ok-xas}
	\end{center}
\end{figure}

The main paper presented XAS spectra at the Ni-$L_3$ edge. Here, in Fig.~\ref{fig:ok-xas} we plot O $K$-XAS spectra measured in both the bulk-sensitive fluorescence yield (FY) and the surface-sensitive total electron yield (TEY) modes. We have measured the four samples studied in the main text, as well as pristine NdNiO$_3$, and the substrate material SrTiO$_3$. From this we see qualitatively little change between all the infinite-layer nickelates for both FY and EY O $K$-XAS spectra. In FY mode there is almost no pre-peak in reduced films comparing to the prominent pre-edge in the NdNiO$_3$ which suggests that the reduction from 113 to 112 phase is almost complete. The TEY mode data show almost the same spectral profile indicating that the pre-peak is unlikely due to the RP defect phase but rather the intrinsic property of the infinite-layer 112 phase.  

\begin{figure}[h]
	\begin{center}
		\includegraphics[width=\textwidth]{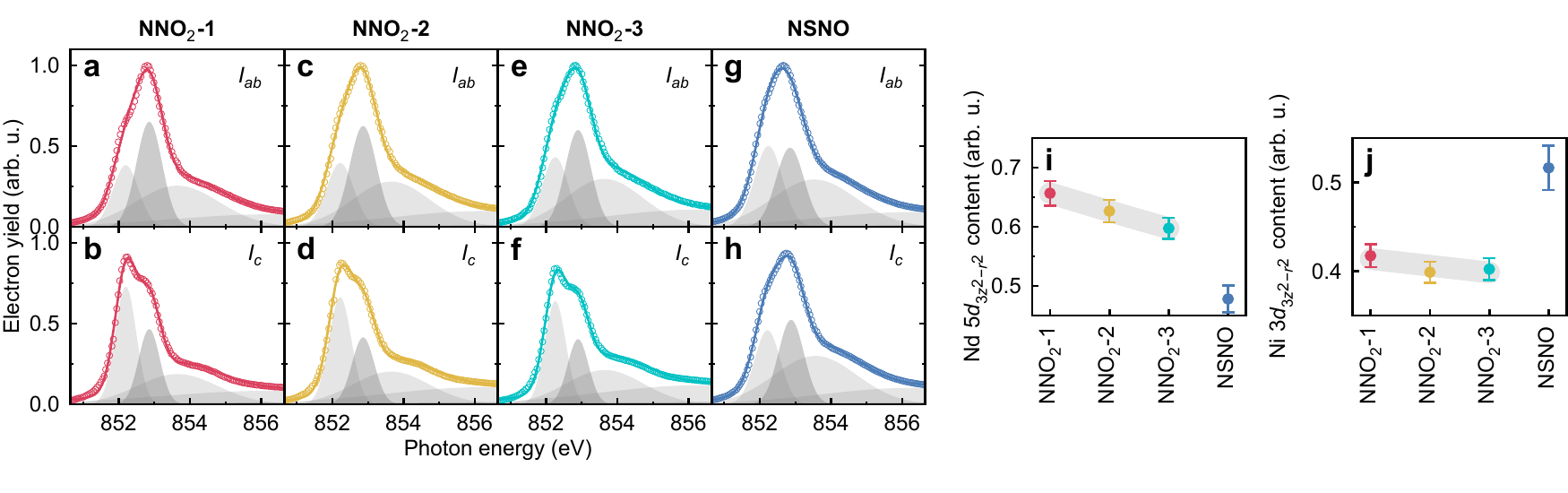}
    	\caption{\textbf{Total electron yield Ni $L_3$ XAS of the parent NdNiO$_2$ and the superconducting Nd$_{0.8}$Sr$_{0.2}$NiO$_2$.} The XAS projected in (out of) the NiO$_2$ planes, I$_{ab}$ (I$_c$), of NNO$_2$-1, NNO$_2$-2, NNO$_2$-3, and NSNO are shown in \textbf{a} (\textbf{b}), \textbf{c} (\textbf{d}), \textbf{e} (\textbf{f}), \textbf{g} (\textbf{h}), respectively. The normalisation and fitting analysis were done in the same way as implemented for the fluorescence yield XAS. The fit components that are visible in this energy range are plotted in different grey colors (See fitting details in Methods). The extracted orbital contents from the surface layers show the same monotonic trend as those from the bulk for NNO$_2$ samples.}
		 \label{fig:e-yield-analysis}
	\end{center}
\end{figure}

In Fig.~\ref{fig:e-yield-analysis} is plotted the same analysis of Ni $L_3$ XAS shown in Fig.~2 in the main text but in TEY mode XAS. The main difference is the absolute change in XAS intensities as a function of samples while the global trend of the orbital content is the same as inferred from the FY mode XAS.

\begin{figure}[h]
	\begin{center}
		\includegraphics[width=0.9\textwidth]{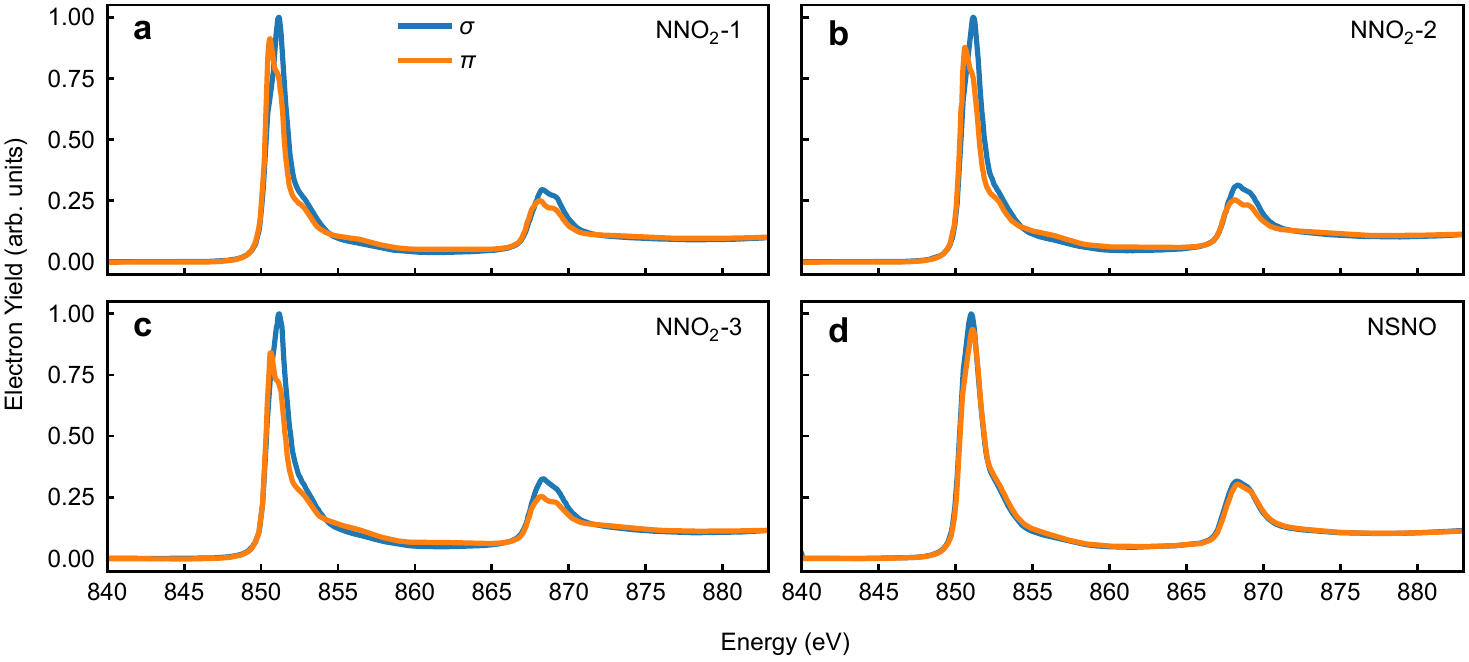}
    	\caption{\textbf{Normalised total electron yield Ni $L_3$-XAS spectra.}}
		 \label{fig:EY_XAS}
	\end{center}
\end{figure}

\begin{figure}[h]
	\begin{center}
		\includegraphics[width=0.9\textwidth]{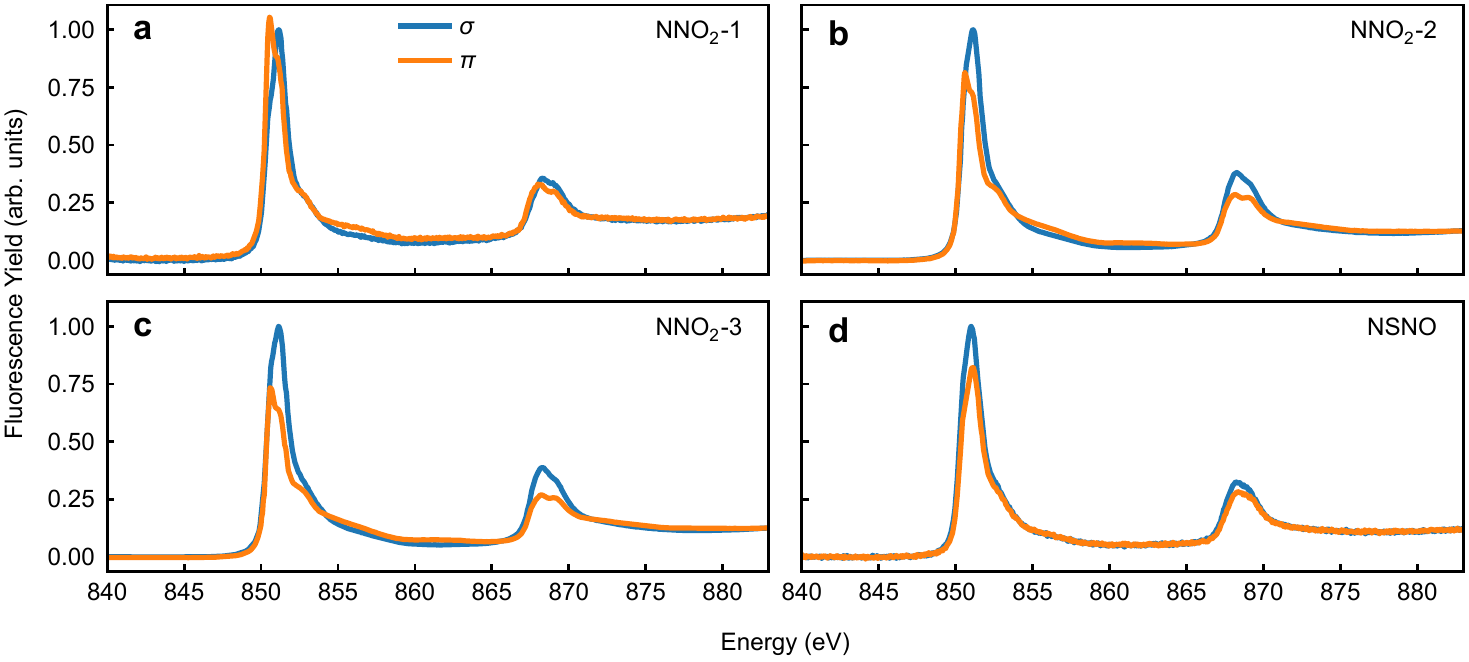}
    	\caption{\textbf{Normalised fluorescence yield Ni $L_3$-XAS spectra.}}
		 \label{fig:FY_XAS}
	\end{center}
\end{figure}

The orbital content analysis in Fig.~2 in the main text and Fig.~\ref{fig:e-yield-analysis} relies heavily on the normalisation. In Figs.~\ref{fig:EY_XAS},~\ref{fig:FY_XAS} we show XAS spectra which have been collected up to the Ni-$L_2$ post edge. All spectra were initially aligned to the pre-edge and then normalised to the post-edge. The $L_2$ absorption peak shows similar dichroism to the $L_3$ edge.

\FloatBarrier

\bibliography{references}